\documentclass[12pt,a4wide]{article}
\usepackage{a4wide,amsmath,amsthm,epsfig,graphicx}
\usepackage{amsmath,amsthm,amssymb}
\usepackage{amsfonts}
\usepackage{graphicx}

\newcommand{\dob}{\mbox{DOB}}

\newcommand{\Qx}{ \mathbb{Q} }
\newcommand{\Ex}{ \mathbb{E} }

\newcommand{\cds}{\mbox{CDS}}

\newcommand{\pcds}{\Pi\mbox{\tiny RCDS}}

\newcommand{\pprcds}{\Pi\mbox{\tiny PRCDS}}

\newcommand{\npv}{\mbox{NPV}}
\newcommand{\defree}{\mbox{\tiny DEFREE}}

\newcommand{\Hhat}{\widehat{H}}
\newcommand{\lgd}{\mbox{L{\tiny GD}}}
\newcommand{\rec}{\mbox{R{\tiny EC}}}

\newtheorem{theorem}{Theorem}[section]
\newtheorem{proposition}[theorem]{Proposition}
\newtheorem{remark}[theorem]{Remark}

\title{\vspace{-1cm } \bf{\small Reduced version in Proceedings of the FEA 2004 Conference at MIT, Cambridge, Massachusetts,
November 8-10, and in: Pykhtin, M. (Editor), Counterparty Credit Risk Modeling: Risk Management, Pricing and Regulation. 
 Risk Books, 2005,  London.} \\ \vspace{1cm} \bf{\Large Credit Default Swap
Calibration and Equity Swap Valuation under Counterparty Risk with
a Tractable
Structural Model}
}
\author{
Damiano Brigo \ \ \ \  Marco Tarenghi \\
Credit Models\\
Banca IMI\\
Corso Matteotti 6 \\
20121 Milano, Italy\\
Posted on SSRN.com on August 24, 2004\\ 
{\tt http://ssrn.com/abstract=581302}\\
\texttt{http://www.damianobrigo.it}
}

\date{First version: March 1, 2004. This Version: March 8, 2005}

\begin{document}

\maketitle \thispagestyle{empty}

\begin{abstract}
In this paper we develop a tractable structural model with
analytical default probabilities depending on some dynamics
parameters, and we show how to calibrate the model using a chosen
number of Credit Default Swap (CDS) market quotes. We essentially
show how to use structural models with a calibration capability
that is typical of the much more tractable credit-spread based
intensity models. We apply the structural model to a concrete
calibration case and observe what happens to the calibrated
dynamics when the CDS-implied credit quality deteriorates as the
firm approaches default. Finally we provide a typical example of a
case where the calibrated structural model can be used for credit
pricing in a much more convenient way than a calibrated reduced
form model: The pricing of counterparty risk in an equity swap.
\end{abstract}

\subsection*{Keywords} Credit Derivatives, Structural Models, Black
Cox Model, Credit Default Swaps, Calibration, Analytical
Tractability, Monte Carlo Simulation, Equity Swaps, Counterparty
Risk, Barrier Options.

\newpage

\tableofcontents

\newpage

\section{Introduction}

Modelling firms default is an important issue, especially in
recent times where the market is experiencing an increasing
interest in credit derivatives trading. Pricing models can be
divided into two main categories: (i) reduced form models and (ii)
structural models.

\emph{Reduced form models} (also called \emph{intensity models}
when a suitable context is possible) describe default by means of
an exogenous jump process; more precisely, the default time is the
first jump time of a Poisson process with deterministic or
stochastic (Cox process) intensity. Here default is not triggered
by basic market observables but has an exogenous component that is
independent of all the default free market information. Monitoring
the default free market does not give complete information on the
default process, and there is no economic rationale behind
default. This family of models is particularly suited to model
credit spreads and in its basic formulation is easy to calibrate
to Credit Default Swap (CDS) data. See references at the beginning
of Chapter 8 in Bielecki and Rutkowski (2001) for a summary of the
literature on intensity models. We cite here  Duffie and Singleton
(1999) and Lando (1998), and Brigo and Alfonsi (2003) as a
reference for explicit calibration of a tractable stochastic
intensity model to CDS data.

\emph{Structural models} are based on the work by Merton (1974),
in which a firm life is linked to its ability to pay back its
debt. Let us suppose that a firm issues a bond to finance its
activities and also that this bond has maturity $T$. At final time
$T$, if the firm is not able to reimburse all the bondholders we
can say that there has been a default event. In this context
default may occur only at final time $T$ and is triggered by the
value of the firm being below the debt level. In a more realistic
and sophisticated structural model (Black and Cox (BC) (1976),
part of the family of \emph{first passage time models}) default
can happen also before maturity $T$. In first passage time models
the default time is the first instant where the firm value hits
from above either a deterministic (possibly time varying) or a
stochastic barrier, ideally associated with safety covenants
forcing the firm to early bankruptcy in case of important credit
deterioration. In this sense the firm value is seen as a generic
asset and these models use the same mathematics of barrier options
pricing models. For a summary of the literature on structural
models, possibly with stochastic interest rates and default
barriers, we refer for example to Chapter 3 of Bielecki and
Rutkowski (2001). It is important to notice that structural models
make some implicit but important assumptions: They assume that the
firm value follows a random process similar to the one used to
describe generic stocks in equity markets, and that it is possible
to observe this value at any time. Therefore, unlike intensity
models, here the default process can be completely monitored based
on default free market information and comes less as a surprise.
However, structural models in their basic formulations and with
standard barriers (Merton, BC) have few parameters in their
dynamics and cannot be calibrated exactly to structured data such
as CDS quotes along different maturities.

In this paper we plan to effectively use a structural model in the
ideal ``territory" of intensity models, i.e. to describe default
probabilities in a way that is rich enough to calibrate CDS
quotes. First of all, in Section~\ref{sec:barrieroptions} we show
how standard barrier options formulas can be used to find these
probabilities, and in particular we investigate the case of time
dependent parameters in the dynamics of the underlying process. In
Section~\ref{sec:calibration} we show how to calibrate the
parameters of the model to market prices of single name CDS's (in
a way similar to the procedure used to calibrate intensities in
reduced form models). In Section~\ref{sec:casestudy} we consider
the concrete case of a firm approaching default, and we show how
our proposed structural model calibration changes as the company
credit quality (as summarized by its CDS quotes) deteriorates in
time. The calibrated model is useful for subsequent pricing of
more sophisticated derivatives depending on default. Indeed, in
Section~\ref{sec:otherproducts} we see an important example of how
our structural model can be used to price some products involving
considerations from both the credit and the equity market. The
product we consider is an equity return swap with counterparty
risk. In this context we have to take care of the correlation
between the counterparty and the underlying, and this is done much
more conveniently in structural models than in intensity models.

Finally, we have considered tractable extensions of the AT1P model
based on introducing scenarios on the value of the firm volatility
and safety barrier. This framework allows to consider situations
where the balance sheets information can be uncertain or hid some
developments. This work and the related SVBAT1P models are
reported in Brigo and Tarenghi (2005), with a calibration case
study on CDS data.

\section{The Black Cox approach and Barrier options formulas}\label{sec:barrieroptions}
The fundamental hypothesis of the model we resume here is that the
underlying process is a Geometric Brownian Motion (GBM), which is
also the kind of process commonly used for equity stocks in the
Black Scholes model.

Classical structural models (Merton, Black Cox) postulate a GBM
(Black and Scholes) lognormal dynamics for the value of the firm
$V$. This lognormality assumption is considered to be acceptable.
Crouhy et al (2000) report that ``this assumption [lognormal V] is
quite robust and, according to KMV's own empirical studies, actual
data conform quite well to this hypothesis.".

In these models the value of the firm $V$ is the sum of the firm
equity value $S$ and of the firm debt value $D$. The firm equity
value $S$, in particular, can be seen as a kind of (vanilla or
barrier-like) option on the value of the firm $V$. Merton
typically assumes a zero-coupon debt at a terminal maturity $T$.
Black Cox assume, besides a possible zero coupon debt, safety
covenants forcing the firm to declare bankruptcy and pay back its
debt with what is left as soon as the value of the firm itself
goes below a ``safety level" barrier. This is what introduces the
barrier option technology in structural models for default.

More in detail, in Merton's model there is a debt maturity
${\bar{T}}$, a debt face value $L$ and the company defaults at
final maturity (and only then) if the value of the firm
$V_{\bar{T}}$ is below the debt $L$ to be paid.

The debt value at time $t<{\bar{T}}$ is thus

\[ D_t = \Ex_t [D(t,{\bar{T}}) \min(V_{\bar{T}},L)] = \Ex_t [D(t,{\bar{T}})[V_{\bar{T}} - (V_{\bar{T}} -
L)^+ ]] =\]\[= \Ex_t [ D(t,{\bar{T}}) [ L - (L-V_{\bar{T}})^+ ]] =
P(t,{\bar{T}}) L - \mbox{Put}(t,{\bar{T}};V_t,L)
\]
where  Put(time, maturity, underlying, strike) is a put option
price, and the stochastic discount factor at time $t$ for maturity
$T$ is denoted by $D(t,T)=B(t)/B(T)$, where $B(t) = \exp(\int_0^t
r_u du)$ denotes the bank-account numeraire, $r$ being the
instantaneous short interest rate. Since we will assume
deterministic interest rates, in our case $D(t,T) = P(t,T)$, the
zero coupon bond price at time $t$ for maturity $T$.

The equity value can be derived as a difference between the value
of the firm and the debt:

\[ S_t = V_t - D_t = V_t - P(t,{\bar{T}}) L +
\mbox{Put}(t,{\bar{T}};V_t,L) = \mbox{Call}(t,{\bar{T}}; V_t,L)\]

so that, as is well known, in Merton's model the equity can be
interpreted as a call option on the value of the firm.

Let us now move to the Black Cox (BC) model. In this model we have
safety covenants in place, in that the firm is forced to reimburse
its debt as soon as its value $V_t$ hits a low enough ``safety
level" $\widehat{H}(t)$. The choice of this safety level is not
easy. Assuming a debt face value of $L$ at final maturity
${\bar{T}}$ as before, an obvious candidate for this ``safety
level" is the final debt present value discounted back at time
$t$, i.e. $L P(t,{\bar{T}})$. However, one may want to cut some
slack to the counterparty, giving it some time to recover even if
the level goes below $L P(t,{\bar{T}})$, and the ``safety level"
can be chosen to be lower than $L P(t,{\bar{T}})$.

We will come back to this issue later on. For the time being we
just assume $V$ to be a geometric Brownian motion.

\begin{equation}\label{undproc}
dV^\ast_t = V^\ast_t\,\mu^\ast_t\,dt+V^\ast_t\,\sigma_t\,dW(t),
 \ \ V^\ast_0 = V_0
\end{equation}
and impose a time dependent safety barrier $H^\ast(t)$. For the
time being let us assume to have constant parameters (i.e.
$\mu^\ast_t = \mu^\ast$ and $\sigma_t = \sigma$) and a constant
barrier ($H^\ast(t) = H^\ast$). We set ourselves in the risk
neutral measure, so that the drift term has the standard form of
interest rate minus a payout ratio  (i.e. $\mu^\ast = r - q^\ast$)
and (\ref{undproc}) is a standard Geometric Brownian Motion. Let
us pretend for a moment that we are just doing derivatives pricing
in an equity market with underlying $V$. In the literature it is
possible to find many analytical formulas for barrier options
pricing such as, for example, knock-in or knock-out options, and
also digital options. One particular case is the down and out
digital option or down and out bond (DOB), that is a contract
paying one unit of currency at maturity ${T}$ if, between the
starting date of the contract and its maturity $T$, the underlying
never touches the barrier $H^\ast$ ($H^\ast<V^\ast_0$) from above.
If we go back to the default interpretation with $H^\ast$ as
safety barrier level and $T<\bar{T}$, this option price is
actually the price of a defaultable zero coupon bond with no
recovery in the structural model framework.
%
%
%
%
Now, if we call $\tau$ the first time instant where the process
hits the barrier from above, the price of this option in the
risk-neutral framework is given by
\begin{equation}\label{dig1}
\dob(0,T) =  \mathbb{E}\{D(0,T)\,\mathbf{1}_{\{\tau
> T \}}\}
\end{equation}
where $\mathbb{E}$ is the risk neutral expectation associated with
the risk neutral measure $\mathbb{Q}$. Under our deterministic
rates framework (or under interest rates that are independent of
the underlying process) we can write
\begin{equation} \label{dig2}
\dob(0,T) = P(0,T)\,\mathbb{E}\{\mathbf{1}_{\{\tau>T\}}\} =
P(0,T)\,\mathbb{Q}\{\tau>T\} .
\end{equation}
The last factor is the (risk-neutral) probability of never
touching the barrier before $T$, also called \emph{survival
probability}. By means of stochastic calculus it is possible to
derive analytically the price of the option by explicitly
computing (\ref{dig2}) (see for example Bielecki and Rutkowski
(2001)):
\begin{eqnarray}\label{survprob1}
\dob(0,T) = P(0,T) \left[\Phi\left(d_1\right) -
\left(\frac{V_0}{H^\ast}\right)^{1-2(r-q^\ast)/(\sigma^2)}\Phi\left(d_2\right)\right]
\end{eqnarray}
with $d_{1,2} = \left(\pm \log
\frac{V_0}{H^\ast}+\left(r-q^\ast-\frac{\sigma^2}{2}\right)T\right)/(\sigma\sqrt{T})$.
By comparing equations (\ref{dig2}) and (\ref{survprob1}) we
obtain
\begin{eqnarray}\label{survprob2}
\mathbb{Q}\{\tau>T\}= \left[\Phi\left(d_1\right) -
\left(\frac{V_0}{H^\ast}\right)^{1-2(r-q^\ast)/(\sigma^2)}\Phi\left(d_2\right)\right].
\end{eqnarray}
Similar formulas can be obtained also for a barrier that is not
necessarily constant in time (or ``flat"), but with a particular
exponential shape (see again Bielecki and Rutkowski (2001) for
more details).

When relaxing the assumption of constant parameters in the
$V^\ast$ dynamics, the situation becomes much more complicated. In
this case, even with a flat barrier, it is not possible to find
closed form pricing formulas. However, some recent work dealing
with option pricing on underlying assets having time dependent
parameters in the dynamics shows that it is possible to find
analytical barrier option prices when the barrier has a particular
curved shape depending partly on the dynamics parameters. See for
example Rapisarda (2003), who builds on the fundamental work of Lo
et al. (2003), where the formulas are expressed in a shape
resembling the classical constant coefficients Black-Scholes
formulas. Again, let us assume an underlying process like
(\ref{undproc}), with $\mu^\ast_t = r_t - q^\ast_t$, where $r$ and
$q^\ast$ are respectively the time varying instantaneous risk free
interest rate and payout ratio, and consider an option with
maturity $T$. Let us take a barrier of the form:
\begin{equation}\label{barrier}
H^\ast(t) =
H\exp\left(-\int_t^T\left(r_s-q^\ast_s-(1+2\beta)\frac{\sigma_s^2}{2}\right)ds\right)
\end{equation}
depending on a parameter $\beta$ and on the constant reference
value $H$. For this special barrier it is possible to obtain exact
formulas for many different barrier options when the underlying is
$V^\ast$. Obviously, if we had to price an option with a given
{\em constant} barrier, the price obtained with these formulas
would not be properly correct, but it would be a good
approximation if we chose the barrier profile $H^\ast(t)$ as close
as possible to the given flat barrier value. It can be shown that
the value of $\beta$ keeping the barrier as flat as possible is
given by
\begin{equation}\label{betaopt}
\beta^* = \frac{\int_{0}^{T}\left(\int_t^{T}\left(
r_s-q^\ast_s-\frac{\sigma_s^2}{2}\right)ds\right)
\left(\int_t^{T}\sigma_s^2ds\right)dt}{\int_{0}^{T}
\left(\int_t^{T}\sigma_s^2ds\right)^2dt}
\end{equation}
and this reduces to $\beta^*=(r-q^\ast-\sigma^2/2)/(\sigma^2)$ in
the case of constant coefficients (where, as expected, $H^\ast(t)
= H$).

Let us go back to the time-varying barrier~(\ref{barrier}).    In
this framework
\begin{equation}\label{digitdyn}
\dob(0,T)  = P(0,T) \cdot\left[\Phi\left(\frac{\log \frac{V_0}{H}
+\int_0^Tv_sds}{\sqrt{\int_0^T\sigma_s^2ds}}\right) -
\left(\frac{H^\ast(0)}{V_0}\right)^{2\beta}\Phi\left(\frac{\log
\frac{H^\ast(0)^2}{V_0 H}
+\int_0^Tv_sds}{\sqrt{\int_0^T\sigma_s^2ds}}\right)\right]
\end{equation}
where $v_t = r_t-q^\ast_t-\frac{\sigma_t^2}{2}$ and, using again
(\ref{dig2}), we find
\begin{equation}\label{survdyn}
\mathbb{Q}\{\tau>T\} = \left[\Phi\left(\frac{\log \frac{V_0}{H}
+\int_0^Tv_sds}{\sqrt{\int_0^T\sigma_s^2ds}}\right) -
\left(\frac{H^\ast(0)}{V_0}\right)^{2\beta}\Phi\left(\frac{\log
\frac{H^\ast(0)^2}{V_0 H}
+\int_0^Tv_sds}{\sqrt{\int_0^T\sigma_s^2ds}}\right)\right]
\end{equation}
where we recall that  $H^\ast(0) =
H\exp\left(-\int_0^T\left(r_s-q^\ast_s-(1+2\beta)\frac{\sigma_s^2}{2}\right)ds\right)$.

In the next section we show how this formula can be used to
calibrate a given structural model to market data, in particular
to CDS's quotes.

\section{Calibration of the structural model to CDS data}\label{sec:calibration}
Since we are dealing with default probabilities of firms, it is
straightforward to think of financial instruments depending on
these probabilities and whose final aim is to protect against the
default event. One of the most representative protection
instruments is the Credit Default Swap (CDS). CDS's are contracts
that have been designed to offer protection against default.
Consider two companies ``A" (the \emph{protection buyer}) and ``B"
(the \emph{protection seller}) who agree on the following.

If a third reference company ``C" (the \emph{reference credit})
defaults at a time $\tau_C \in (T_a,T_b]$, ``B" pays to ``A" at
time $\tau=\tau_C$ itself a certain ``protection" cash amount
$\lgd$ (Loss Given the Default of ``C" ), supposed to be
deterministic in the present paper. This cash amount is a {\em
protection} for ``A" in case ``C" defaults. A typical stylized
case occurs when ``A" has bought a corporate bond issued from ``C"
and is waiting for the coupons and final notional payment from
this bond: If ``C" defaults before the corporate bond maturity,
``A" does not receive such payments. ``A" then goes to ``B" and
buys some protection against this risk, asking ``B" a payment that
roughly amounts to the bond notional in case ``C" defaults.

Typically $\lgd$ is equal to a notional amount, or to a notional
amount minus a recovery rate. We denote the recovery rate by
``$\rec$".

In exchange for this protection, company ``A" agrees to pay
periodically to ``B" a fixed ``running" amount $R$, at a set of
times $\{T_{a+1},\ldots,T_b\}$,  $\alpha_i = T_{i}-T_{i-1}$,
$T_0=0$. These payments constitute the ``premium leg" of the CDS
(as opposed to the $\lgd$ payment, which is termed the
``protection leg"), and $R$ is fixed in advance at time $0$; the
premium payments go on up to default time $\tau$ if this occurs
before maturity $T_b$, or until maturity $T_b$ if no default
occurs.

\[ \begin{array}{ccccc} \mbox{``B"} & \rightarrow & \mbox{ protection } \lgd \mbox{ at default $\tau_C$ if $T_a< \tau_C \le T_b$} & \rightarrow & \mbox{``A"} \\
%
\mbox{``B"} & \leftarrow  & \mbox{ rate } R \mbox{ at }
T_{a+1},\ldots,T_b \mbox{ or until default } \tau_C & \leftarrow &
\mbox{``A"}
\end{array} \]

Formally, we may write the RCDS (``R" stands for running)
discounted value at time $t$ seen from ``A" as
\begin{eqnarray}\label{discountedpayoffcds}
\nonumber \pcds_{a,b}(t) := - D(t,\tau) (\tau-T_{\beta(\tau)-1}) R
\mathbf{1}_{\{T_a < \tau < T_b \} }
 -   \sum_{i=a+1}^b D(t,T_i) \alpha_i R \mathbf{1}_{\{\tau \ge T_i\}
 } \\
 + \mathbf{1}_{\{T_a < \tau \le T_b \} }D(t,\tau) \ \lgd
\end{eqnarray}
where $t\in [T_{\beta(t)-1},T_{\beta(t)})$, i.e. $T_{\beta(t)}$ is
the first date among the $T_i$'s that follows $t$, and where
$\alpha_i$ is the year fraction between $T_{i-1}$ and $T_i$.

Sometimes a slightly different payoff is considered for RCDS
contracts. Instead of considering the exact default time $\tau$,
the protection payment $\lgd$ is postponed to the first time $T_i$
following default, i.e. to $T_{\beta(\tau)}$. If the grid is three
or six months spaced, this postponement consists in a few months
at worst. With this formulation, the CDS discounted payoff can be
written as
\begin{eqnarray}\label{cdspostponedpayoff}
\pprcds_{a,b}(t) := - \sum_{i=a+1}^b D(t,T_i) \alpha_i R
\mathbf{1}_{\{\tau \ge T_i\} }
 + \sum_{i=a+1}^b \mathbf{1}_{\{T_{i-1} < \tau \le T_i \} }D(t,T_i) \
 \lgd,
\end{eqnarray}
which we term ``Postponed Running CDS" (PRCDS) discounted payoff.
Compare with the earlier discounted
payout~(\ref{discountedpayoffcds}) where the protection payment
occurs exactly at $\tau$: The advantage of the postponed
protection payment is that no accrued-interest term in
$(\tau-T_{\beta(\tau)-1})$ is necessary, and also that all
payments occur at the canonical grid of the $T_i$'s. The postponed
payout is better for deriving market models of CDS rates dynamics
and for relating CDS's to floaters, see for example Brigo~(2004,
2004b). When we write simply ``CDS" we refer to the RCDS case.

Let us consider again the basic RCDS: The pricing formula for this
payoff depends on the assumptions on the interest rates dynamics
and on the default time $\tau$. Let $\mathcal{F}_t$ denote the
basic filtration without default, typically representing the
information flow of interest rates and possibly other default-free
market quantities (and also intensities in the case of reduced
form models), and $\mathcal{G}_t = \mathcal{F}_t\vee
\sigma\left(\{\tau<u\},u\leq t\right)$ the extended filtration
including explicit default information. In our current
``structural model" framework with deterministic default barrier
the two sigma-algebras coincide by construction, i.e.
$\mathcal{G}_t = \mathcal{F}_t$, because here the default is
completely driven by default-free market information. This is not
the case with intensity models, where the default is governed by
an external random variable and $\mathcal{F}_t$ is strictly
included in $\mathcal{G}_t$, i.e. $\mathcal{F}_t \subset
\mathcal{G}_t$.

We denote by $\cds(t,[T_{a+1},\ldots,T_b],T_a,T_b, R, \lgd)$ the
price at time $t$ of the above standard running CDS. At times some
terms are omitted, such as for example the list of payment dates
$[T_{a+1},\ldots,T_b]$. In general we can compute the CDS price
according to risk-neutral valuation (see for example Bielecki and
Rutkowski (2001)):
\begin{equation}\label{priceCDStheo}
\cds(t,T_a,T_b,R,\lgd) =
\mathbb{E}\{\pcds_{a,b}(t)|\mathcal{G}_t\} =
\mathbb{E}\{\pcds_{a,b}(t)|\mathcal{F}_t\} =:
\mathbb{E}_t\{\pcds_{a,b}(t)\}
\end{equation}
in our structural model setup. A CDS is quoted through its ``fair"
$R$, in that the rate $R$ that is quoted by the market at time $t$
satisfies $\cds(t,T_a,T_b,R,\lgd) = 0$. Let us assume, for
simplicity, deterministic interest rates; then we have
\begin{eqnarray}\label{priceCDS}
CDS(t,T_a,T_b,R,\lgd) :=
-R\,\mathbb{E}_t\{P(t,\tau)(\tau-T_{\beta(\tau)-1})\mathbf{1}
_{\{T_a<\tau<T_b\}}\} \nonumber \\  -\sum_{i=a+1}^bP(t,T_i)
\alpha_iR\,\mathbb{E}_t\{\mathbf{1}_{\{\tau\geq T_i\}}\} +
\lgd\,\mathbb{E}_t\{\mathbf{1}_{\{T_a<\tau\leq T_b\}}P(t,\tau)\}.
\end{eqnarray}
It is clear that the fair rate $R$ strongly depends on the default
probabilities. The idea is to use quoted values of these fair
$R$'s with different maturities to derive the default
probabilities assessed by the market.

While in simple intensity models the survival probabilities can be
interpreted as discount factors (with credit spreads as
discounting rates), and as such can be easily stripped from CDS's
or corporate bonds, in structural models the situation is much
more complicated.
%
%
In fact, here, it is not possible to find a simple ``credit
spread" formulation for $d\mathbb{Q}\{ \tau > t\}$ starting from
(\ref{survdyn}). Moreover we have to pay attention to one
fundamental aspect of the model. The barrier triggering default
depends on the maturity chosen for a particular instrument, i.e.
$H^\ast(t) = H^\ast_T(t)$ has a parametric dependence on $T$. But
then, if we plan to use the model for instruments with different
maturities, it is necessary to impose a consistency condition,
that is
\begin{equation}\label{cohercond}
H^\ast_{T_1}(t) = H^\ast_{T_2}(t)
\end{equation}
for $t\leq T_1 < T_2$ for every pair of maturities $T_1$ and
$T_2$. With our choice of $H^\ast(t)$ above we have easily that
this condition implies, when enforced for all possible $T_1, T_2$:
%
%
\begin{equation}\label{coherpunct}
r_t-q^\ast_t-(1+2\beta)\frac{\sigma_t^2}{2} = 0, \ \ \mbox{or} \ \
q^\ast_t =  r_t-(1+2\beta)\frac{\sigma_t^2}{2}.
\end{equation}
We notice that if this condition is satisfied, we have that the
barrier $H^\ast$, defined in terms of $q^\ast$, obviously flattens
to the constant value $H$. Thus {\em consistency of the curved
barrier for all maturities induces a flat barrier}. In general,
however, we expect (\ref{coherpunct}) not to hold if $V^\ast$ is
the value of the firm, since $r$, $q^\ast$ and $\sigma$ are given
to us exogenously and $\beta$ is just constant. However, we can
still manage to preserve analytical tractability as follows.
%
%
%
Let us assume from now on that the real risk neutral dynamics of
the firm value is a process given by
\begin{equation}\label{realund}
dV_t = V_t\,(r_t-q_t)\,dt+V_t\,\sigma_t\,dW(t)
\end{equation}
%
where $q_t$ is the true payout ratio and $q_t^*$ is defined by
(\ref{coherpunct}). The ``system" $(V^\ast,H^\ast)$ with said
$q^\ast$ is both tractable and satisfying the consistency
condition. Our problem is that $V^\ast$ is not the real firm
value, so that the system $(V^\ast,H^\ast)$ is not good for
modeling default. However,  if we define
\begin{equation}\label{realbarrier}
\Hhat(t) = H\exp\left(-\int_0^t(q_s-q_s^*)ds\right)
\end{equation}
by integrating $V^\ast$ and $V$'s equations it is easy to show
that the first time $V^\ast_t$ hits $H^\ast(t)=H$ is the same as
the first time the real process $V_t$ hits $\Hhat(t)$. Therefore,
default probabilities computed with the tractable model
$V^\ast,H^\ast$ are the same as default probabilities for the
``true" model $V,\Hhat$. We can compute quantities and perform our
calibration with the former model and consider the latter as the
real model. We can thus state the following

\begin{proposition} {\bf (Analytically-Tractable
First Passage (AT1P) Model)} Assume the risk neutral dynamics for
the value of the firm $V$ is characterized by a risk free rate
$r_t$, a payout ratio $q_t$ and an instantaneous volatility
$\sigma_t$, according to equation~(\ref{realund}), i.e.
\[ dV_t = V_t\,(r_t-q_t)\,dt+V_t\,\sigma_t\,dW(t)\] and assume a
default barrier $\Hhat(t)$ of the form given
in~(\ref{realbarrier}) with $q^\ast$ given as
in~(\ref{coherpunct}), i.e.
\[ \Hhat(t) =H\exp\left(-\int_0^t\left(q_s-r_s + (1+2\beta)
\frac{\sigma_s^2}{2}\right)ds\right)\] and let $\tau$ be defined
as the first time where $V$ hits $\Hhat$ from above, starting from
$V_0>H$,
\[ \tau = \inf\{ t \ge 0: V_t \le \Hhat(t)\}. \]
Then the survival probability is given analytically by
\begin{eqnarray}\label{survcoher}
\mathbb{Q}\{\tau>T\} = \left[\Phi\left(\frac{\log \frac{V_0}{H}
+\beta\int_0^T \sigma_s^2 ds}{\sqrt{\int_0^T\sigma_s^2ds}}\right)-
\left(\frac{H}{V_0}\right)^{2\beta}\Phi\left(\frac{\log
\frac{H}{V_0} +\beta \int_0^T \sigma_s^2
ds}{\sqrt{\int_0^T\sigma_s^2ds}}\right)\right].
\end{eqnarray}
\end{proposition}
Formula~(\ref{survcoher}) is easily obtained by substituting the
consistency condition (\ref{coherpunct}) in formula
(\ref{survdyn}) for the survival probability. From our earlier
definitions, straightforward computations lead to the price at
initial time $0$ of a CDS, under deterministic interest rates, as
\begin{eqnarray}\label{cdsformulaat1p}
\cds_{a,b}(0,R,\lgd) = R \int_{T_a}^{T_b} P(0,t)
(t-T_{\beta(t)-1}) d \Qx(\tau > t)\\ \nonumber - R \sum_{i=a+1}^b
P(0,T_i) \alpha_i \Qx(\tau \ge T_i)
 - \lgd \int_{T_a}^{T_b} P(0,t) d \Qx(\tau
> t)
\end{eqnarray}
so that if one has a formula for the curve of survival
probabilities $t \mapsto \Qx(\tau
> t)$, as in our AT1P structural model, one also has a formula for CDS.

Notice an important feature of Formula~(\ref{survcoher}) and then
(\ref{cdsformulaat1p}): Survival (and default) probabilities and
CDS values only depend on the ratio between $V$ and $H$, and not
on $V$ and $H$ separately. This means that, as far as default
probabilities are concerned, a precise estimation of $V$ is not
needed. We can express default probabilities in relative terms,
i.e. in terms of $V/H$. In other terms, we can imagine to re-scale
$V$ (considering $V/V_0$ so that the initial condition reads
$V_0/V_0=1$) and also the barrier parameter $H$ (taking $H/V_0$, a
number smaller than one as barrier).

Formula~(\ref{survcoher}) can be used to fit the model parameters
to market data. However, the only parameter left that can account
for time dependence is the volatility. If we use exogenous
volatility (deduced perhaps from historical or implied equity
volatility) we are left with no freedom. However, we may infer the
first year volatility $\sigma(0\div 1y):=\{\sigma(t): t\in
[0,1]\}$ from equity data and use $H$ as a first fitting
parameter, and then use the remaining later volatilities
$\sigma(1y\div 2y),\sigma(2y\div 3y)$ etc as further fitting
parameters. Therefore $\sigma(2y\div \cdot)$ will be determined by
credit quality as implied by CDS data rather than by equity data.
To sum up, we can choose piecewise constant volatility, and look
for those volatility values after the first year that make the
quoted CDS's fair when inserting in their premium legs the market
quoted $R$'s. In this way we find as many volatilities as many
CDS's we consider minus one. In this first case $H$ is determined
by CDS's and we explain below how exactly we find this $H$ (credit
spread method).

Alternatively, if we aim at creating a one to one correspondence
to volatility parameters and CDS quotes, we can exogenously choose
the value $H$ (for example we will see below the
protection/excursion method, but judgemental analysis or trial and
error are often needed) and $\beta$, leaving all the unknown
information in the calibration of the volatility. If we do so, we
find exactly one volatility parameter for each CDS maturity,
including the first one.

In general the above CDS calibration procedures are justified by
the fact that in the end we are not interested in estimating the
real process of the firm value underlying the contract, but only
in reproducing risk neutral default probabilities with a model
that makes sense also economically. While it is important that the
underlying processes have an economic interpretation, we are not
interested in sharply estimating them or the capital structure of
the firm, but rather we appreciate the structural model
interpretation as a tool for assessing the realism of the outputs
of calibrations, and as an instrument to check economic
consequences and possible diagnostics.

Finally, in case we still have a preferred terminal maturity
${\bar{T}}$ for the debt, and an  indication of the final debt at
maturity, which we call $L$, it makes sense  to impose that our
barrier $\widehat{H}$ be always below the present value of the
final debt, i.e. $P(t,{\bar{T}}) L$. This condition amounts to
assuming that we are cutting some slack to the firm by allowing it
to go somehow below the debt present value before forcing it to
declare bankruptcy. How much below can be decided by means of $H$
and $\beta$. In detail, our condition reads

\[H\exp\left(-\int_0^t\left(q_s-r_s + (1+2\beta)
\frac{\sigma_s^2}{2}\right)ds\right) < \exp\left(-\int_t^{\bar{T}}
r_s ds\right) L . \]

We can easily rewrite this condition as

\[ L > H \exp\left[-\int_0^t\left(q_s+ (1+2\beta)
\frac{\sigma_s^2}{2}\right)ds + \int_0^{\bar{T}} r_s ds \right]\]

A sufficient condition for this, in case the round brackets term
is positive, is

\[ H \le L P(0,\bar{T}),\]

i.e. the final debt initial present value has to larger than the
barrier parameter $H$. This means that in all our use of the model
with positive round brackets one may presume a final debt $L$ at a
preferred maturity (typically larger than ten years, since this is
the largest CDS maturity) satisfying this condition.

We are still in need to connect equity and firm value if we aim at
deriving part of the value of the firm volatility from equity
data. Since default is enforced as soon as $V$ hits $\widehat{H}$
at time $\tau$ if before ${\bar{T}}$, or is given at ${\bar{T}}$
if $V_{\bar{T}}$ is below $L$, the debt value at time $t$ would be
\[ D_t = \Ex_t [ D(t,\tau) \widehat{H}(\tau) \mathbf{1}_{\{\tau < {\bar{T}}\}} ]
+ \Ex_t [ D(t,{\bar{T}}) \min(L,V_{\bar{T}})  \mathbf{1}_{\{\tau
\ge {\bar{T}}\}} ]\] which can be computed as a function of $V_t$
with computations similar to~(\ref{survcoher}) (with time $t$
replacing time $0$). In turn, we would have the equity value as
$S_t = V_t - D_t = V_t - D_t(V_t)$ from which we could derive an
approximation for $V$'s (unknown) volatility in terms of $S$'s
(known, be it historical or implied) volatility through Ito's
formula and some approximations.

However, we will not pursue this strategy but simply take the
equity volatility itself as a proxy for the order of magnitude of
the firm value volatility. As a matter of fact, our preferred
approach will be to let CDS data select most (and in some cases
all) values of the firm volatilities.

\subsection{First Numerical Example}
In this section we present some results of the calibration
performed with the structural model. We consider CDS contracts
having the telecoms sector Vodafone company
%
%
as underlying with recovery rate $\rec=40\%$ ($\lgd=0.6$). In
Table \ref{VodSpread} we report the maturities $T_b$ of the
contracts and the corresponding ``mid" CDS rates $R_{0,b}^{\tiny
\mbox{MID}}(0)$ (quarterly paid) on the date of March 10th, 2004,
in basis points ($1bp = 10^{-4}$). We take $T_a=0$ in all cases.

\begin{table}[h!]
\begin{center}
\begin{tabular}{|c|c|}
\hline Maturity $T_b$ & Rate $R_{0,b}^{\tiny \mbox{MID}}(0)$ (bps)\\\hline March 21st, 2005 & 21.5 \\ March 20th, 2007 & 33.0 \\ March 20th, 2009 & 43.0 \\
March 21st, 2011 & 49.0\\ March 20th, 2014 & 61.0
\\ \hline
\end{tabular}
\caption{\small Maturities of Vodafone's CDS's with their
corresponding rates on March 10, 2004.} \label{VodSpread}
\end{center}
\end{table}

In Tables \ref{volaTab} and \ref{intTab} we present the results of
the calibration performed with the structural model and, as a
comparison, of the calibration performed with a deterministic
intensity (credit spread) model (using piecewise linear
intensity). In this first example the parameters used for the
structural model have been selected based on qualitative
considerations, and are $\beta = 0.5$  and $H/V_0=0.5$. The
$\sigma$'s have been found by calibration to CDS quotes. Below we
report both the values of the calibrated parameters in the two
models (volatilities and intensities) and the related survival
probabilities.

\begin{table}[h!]
\begin{center}
\begin{tabular}{|c|c|c|}
\hline Maturity $T_b$ & Volatility $\sigma(T_{b-1}\div T_b)$ & Survival Prob. $\mathbb{Q}(\tau > T_b)$ \\ \hline March 10th, 2004 &   24.343\% &  100.000\% \\

 March 21st, 2005 &   24.343\% &   99.625\% \\

 March 20th, 2007 &   12.664\% &   98.315\% \\

 March 20th, 2009 &   12.766\% &   96.352\% \\

 March 21st, 2011 &   12.659\% &   94.204\% \\

 March 20th, 2014 &   15.271\% &   89.645\% \\
 \hline
\end{tabular}
\end{center}
\caption{\small Calibrated piecewise constant volatilities nodes
and subsequent survival probabilities with the structural model.
In the first column there are the maturities of the contracts (the
first entry is the date where the calibration is performed. All
$\sigma$'s have been obtained from CDS quotes, $H$ and $\beta$
being fixed exogenously). }\label{volaTab}
\end{table}

\begin{table}[h!t]
\begin{center}
\begin{tabular}{|c|c|c|}
\hline Maturity & Intensity & Survival Prob. $\mathbb{Q}(\tau > T_b)$ \\ \hline March 10th, 2004 &    0.357\% &  100.000\% \\

 March 21st, 2005 &    0.357\% &   99.627\% \\

 March 20th, 2007 &    0.952\% &   98.316\% \\

 March 20th, 2009 &    1.033\% &   96.355\% \\

 March 21st, 2011 &    1.189\% &   94.206\% \\

 March 20th, 2014 &    2.104\% &   89.604\% \\
 \hline
\end{tabular}
\end{center}
\caption{\small Intensity model. In the second column we present
the nodes of the piecewise linear calibrated intensities  and in
the last column the corresponding survival
probabilities.}\label{intTab}
\end{table}

\begin{figure}[!h]
\begin{center}
\includegraphics[angle=270,width=0.8\textwidth]{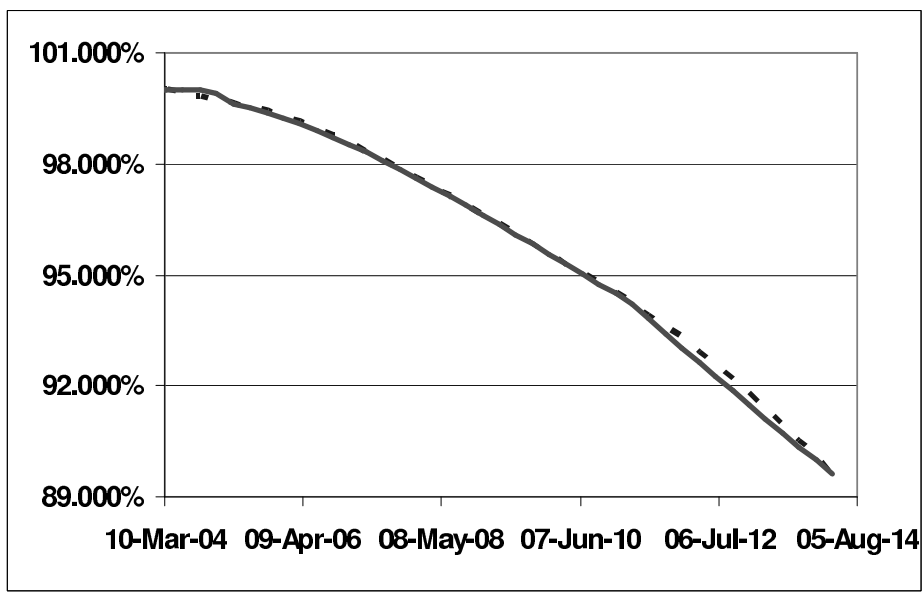}
\end{center}
\caption{\small Survival probabilities $T\mapsto \Qx \{\tau
> T\}$ comparison between the
CDS calibration with the structural model and the CDS calibration
with the intensity model. The dotted line corresponds to survival
probabilities computed with the intensity model, the continuous
line to survival probabilities computed with the structural
model.}\label{survIntStr}
\end{figure}

As a further comparison, we plot the behavior of the survival
probabilities $T\mapsto \Qx \{\tau
> T\}$ resulting from CDS calibration in both cases (Figure \ref{survIntStr}). It is
clear that the survival probability found with the two different
models is nearly the same, which indirectly says that CDS are
actually instruments that efficiently translate default
probabilities into prices, since the default probabilities depend
little on the chosen model, as long as this is consistent with the
same CDS quotes. This is somehow obvious when keeping in
mind~(\ref{cdsformulaat1p}).

A difference is in the first period, i.e. during the life of the
first CDS, where the probability computed with the structural
model is slightly above the other one, given the less sudden
nature of the default in structural models in general.  This is
illustrated in Figure \ref{survIntStr}. This leads us to the
following

\begin{remark} {\bf (Short term credit spreads)}.
It is often said that structural models imply unrealistic
short-term credit spreads. However, if ``short-term" is meant as a
realistic short maturity, our model does not suffer from this
drawback, since it can calibrate any realistic 6m or hypothetical
3m CDS quote. If one is not happy with the default probability
between 0 and 1y, it suffices to calibrate a shorter term CDS, and
the model can do this exactly. What is more, we will see below a
case with very high default probability where only the structural
model is able to calibrate the CDS quotes, the intensity model
giving negative intensities as outputs.
\end{remark}

\begin{figure}[!h]
\begin{center}
\includegraphics[angle=270,width=0.8\textwidth]{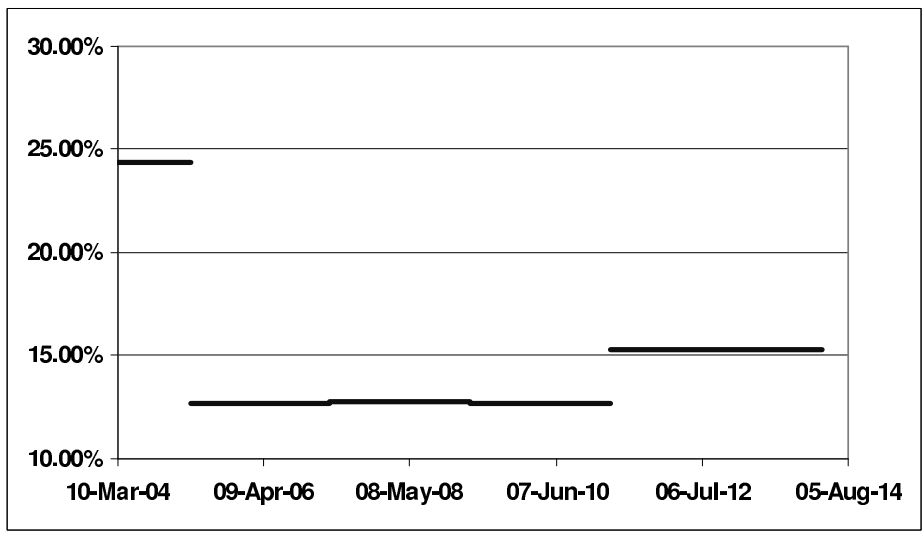}
\end{center}
\caption{\small Volatility term structure $T \mapsto \sigma(T)$
obtained from the calibration. The corresponding data are
presented in Table \ref{volaTab}.}\label{volaTS}
\end{figure}

In Figure~\ref{volaTS} we plot the term structure of the
calibrated volatility (see also Table~\ref{volaTab}). Sometimes
the calibration yields values that are not plausible as volatility
levels, but these volatilities are to be taken as default
probabilities fitting parameters with an economic interpretation
rather than directly as firm value volatilities. In any case,
consistency with CDS quotes for relative value pricing is an
important aspect we need to incorporate in any relative value
pricing model. Also, embedding CDS information in a structural
model framework allows us to check a posteriori the realism of the
calibrated quotes and of the model, so that we may play with our
degrees of freedom (for example in $\beta$ and $H$) to obtain more
realistic calibrations.

We can double check the validity of our results above by means of
a Monte Carlo simulation. We can simulate numerically the process
$V^\ast$ and check whether it crosses the default barrier $H^\ast$
or not: In both cases we can compute the (discounted) payoff of
the CDS for that particular path. We repeat this procedure $N$
times and find the expected value of the payoff as a sample mean
from the simulated scenarios. If, for each maturity $T_b$ we use
the rate $R_{0,b}(0)$ given by the market, then we should obtain
CDS's with zero values.

First of all we derive the right dynamics for the underlying: We
use the process~(\ref{undproc}) using the consistency
condition~(\ref{coherpunct}), obtaining
\begin{equation}\label{MCundproc}
dV^\ast_t =
V^\ast_t\,(1+2\beta)\frac{\sigma_t^2}{2}dt+V^\ast_t\,\sigma_t\,dW(t).
\end{equation}
We know that in general simulations of processes with barriers
require a very large number of paths and a very small time step.
To avoid this problem we applied the Brownian Bridge method (see
for example Metwally and Atiya (2002)), which allows us to use a
larger time step (actually we used an interval of about five
days), thus reducing significantly the simulation time without
losing precision on the reproduced default probabilities.

In Table \ref{MCcheck} we present the result of the Monte Carlo
simulation. We used the same data considered at the beginning of
the section, and the rates $R$ given in Table~\ref{VodSpread}. We
used $N=250000$ scenarios, finding that the values of the CDS's
simulated according to the analytically calibrated dynamics are
practically zero as should be (zero is inside the small standard
error MC window).

\begin{table}[h!]
\begin{center}
\begin{tabular}{|c|c|c|}
\hline Maturity & Simulated Payoff & MC std. dev.\\
\hline
March 21st, 2005 & 0.2 & 0.7 \\
March 20th, 2007 & 0.7 & 1.5 \\
March 20th, 2009 & -0.9 & 2.1 \\
March 21st, 2011 & -0.8 & 2.5 \\
March 20th, 2014 & -0.1 & 3.1 \\
\hline
\end{tabular}
\end{center}
\caption{\small Simulated payoffs of the previously considered
CDS's in basis points. The errors of the Monte Carlo simulations
are reported as well. The number of scenarios is
$N=250000$.}\label{MCcheck}
\end{table}

\subsection{Finding $H$: Credit spread method and protection/excursion analogy method}
Not all possible values of the parameters $H$ and $\beta$ are good
for the calibration. The parameter $\beta$ can be used to shape
the safety covenants barrier $\Hhat$ once $q,\sigma$ and $r$ are
given. And if we calibrate the $\sigma$'s, the initial choice of
$\beta$ will have repercussions on the calibration output
patterns. In general for a given $\beta$, if $H$ is too high (or
too low) the calibration may not succeed in finding all the
volatilities. A way to choose a valid reference value $H$ having
also a more direct link with market information is the following.
Suppose we know the equity volatility for one-year maturity (for
example an implied volatility taken from quoted options) and
suppose we set the structural model's $\sigma$ up to one year to a
constant value given by said equity volatility. Then the
probability of not defaulting before $T$ ($=1$ year) is given by
(see (\ref{survcoher}))
\begin{equation*}
\mathbb{Q}\{\tau>T\} = \left[\Phi\left(\frac{\log \frac{1}{H}
+\beta\sigma^2 T}{\sigma\sqrt{T}}\right) -
H^{2\beta}\Phi\left(\frac{\log H +\beta \sigma^2
T}{\sigma\sqrt{T}}\right)\right]=\psi(H;\beta,T,\sigma)
\end{equation*}
(we have re-scaled $V$ to one, i.e. $V_0=1$, so that the found $H$
will have to be multiplied by $V_0$) and depends on $H$. Being all
other parameters known, $\psi$ is monotonically decreasing in $H$.
But if we adopt a deterministic intensity model for one moment,
the market provides us with
\begin{equation*}
\mathbb{Q}\{ \tau
> T \} = \exp\left(-\int_0^T \lambda_{CDS}(s) ds\right)
\end{equation*}
where $\lambda_{CDS}$ are the deterministic intensities that have
been stripped from CDS data. To impose a reasonable barrier $H$ we
 solve
\begin{equation*}\label{barriercond}
\exp\left(-\int_0^T \lambda_{CDS}(s) ds\right) =
\psi(H;\beta,T,\sigma)
\end{equation*}
in $H$. Since $\psi$ is monotonic, this should be easily dealt
with numerically. We may call this value the ``credit spread based
$H$", since $H$ comes from a default probability based on credit
spreads stripped by CDS's. Using this $H$, we calibrate the
structural model volatilities to the CDS quotes, including the
first one. Since the intensity-based 1y default probability
extracted from CDS and the CDS quote itself contain similar
information but are not exactly equivalent, we will find a
$\sigma(0\div 1y)$ that is close to our initial one-year $\sigma$
selected from the equity market, but not exactly the same. This is
due for example to the first CDS having quarterly payments on
whose dates survival probabilities depend on the chosen model.
Even if two models agree on the one-year survival probability,
they are not necessarily agreeing on the say semiannual default
probability. And if a payoff involves said default probability the
two models may give slightly different values even if they agree
on the one-year default probability. This is why for example in
Table~\ref{output10sep} we find $\sigma(0\div 1y) = 5.012\%$ when
the equity volatility is $5\%$.

A different possibility for choosing $H$ is the following. We may
compare the range of protection $\lgd$ offered with the CDS to a
rough covering of the excursion $V-H$ characterizing default in
the structural model. If we consider then $\lgd V_0$ as the
protection (on the notional $V_0$ associated with the initial
value of the firm) and set it equal to $V_0 - H$ we find
\[ \lgd = 1-H/V_0, \ \ \mbox{or} \ \ \ H/V_0 = \rec, \]
so that we have an immediate candidate for $H$. We term this value
the ``excursion/protection analogy $H$". However, if we resort to
this value we may have numerical problems in stripping
volatilities and moreover we cannot impose the one-year volatility
from the equity market any longer, since we have already used the
degree of freedom provided by $H$. Indeed the calibration proceeds
in the same way as before, but problems may arise in low recovery
cases. Low values for recovery rates imply low barriers, which in
turn lead to high values for stripped volatilities, even much
larger than $100\%$. Under these conditions there are times when
the solver is not able to retrieve a feasible solution for the
calibrated volatilities. A way to bypass this obstacle is given by
the parameter $\beta$: its main role is to vary the steepness of
the safety covenant barrier, as is clear from (\ref{realbarrier}),
and this fact has a relevant impact also on the term structure of
volatilities resulting from the calibration, as we explain more in
detail in the following section. As a matter of fact, it is
possible to indirectly act on the volatilities by varying $\beta$,
looking for those values of $\beta$ returning reasonable values of
$\sigma$ from the calibration, while keeping an eye on the realism
of the resulting safety barrier.

\section{Further Model Specification and a Case Study: the Parmalat Crisis}\label{sec:casestudy}
We now apply the model to a concrete case. We are going to see
what happens, in terms of model parameters, when a firm approaches
default. In particular we consider the case of Parmalat
Finanziaria SpA, an important Italian company whose main activity
is the production of milk and other alimentary goods, and also
present in a large variety of other sectors. In the past years
this company and the related stock have been very important for
the Italian economy; in fact in 2003 the Parmalat stock was
included in the main Italian stock index (Mib 30). After some
signal worrying the investors, at the end of 2003 the Italian
control commission (Consob) began to analyze deeply the balance
sheets of the company, and as the investigation proceeded
%
the company entered a deep crisis.

\begin{figure}[!h]
\begin{center}
\includegraphics[angle=270,width=0.8\textwidth]{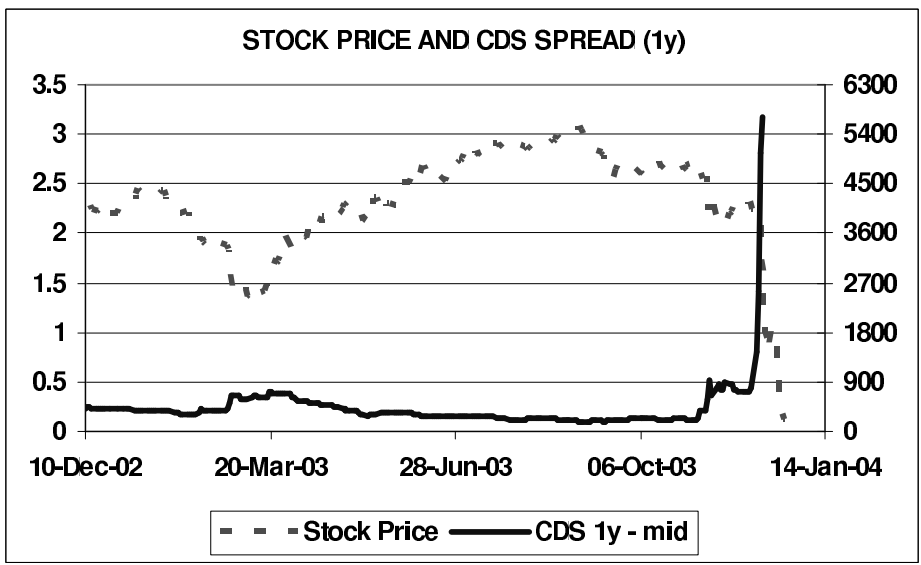}
\end{center}
\caption{\small Historical behavior of the Parmalat stock $S_t$
(in euros) and of the 1y-CDS $R(t)$ (in bps). The dotted line
corresponds to the stock price (left scale) while the continuous
line to the CDS rate (right scale).}\label{prices}
\end{figure}

Here we plan to analyze how the structural model behaves in such a
situation, as the credit quality implied by CDS quotes
deteriorates in time. First of all, in Figure~\ref{prices} we
present the graph of the stock price and the graph of the (1y-)
CDS rate $R$. It is immediate to see that there exists a sort of
negative correlation between the two, as expected: As the equity
value $S$ increases, the CDS premium $R$ (expressing a protection
cost) decreases. From the historical data, we see that the stock
price experienced a deep fall on December 11th. So we consider
some particular dates before that day, some in ``standard regime
conditions" and also some dates in ``crisis conditions", to see
how the CDS calibration changes with market conditions.

In Table \ref{PARhistory} we report the considered dates, with the
related stock prices, CDS rates (in basis points), and the
recovery rates. We also add the volatilities: These are historical
volatilities obtained by the last month of data (as records of
implied volatility were not available). These volatilities have
been sometimes approximated in excess, to take into account the
fact that in crisis periods the stock becomes more volatile than
in the previous period, an effect that is not modelled by
historical volatility.

\begin{table}[h!]
\begin{center}
\begin{tabular}{|c|c|c|c|c|}
\hline Date & Sep 10th & Nov 28th & Dec 8th & Dec 10th\\
\hline R (1y) & 192.5 & 725 & 1450 & 5050\\
R (3y) & 215 & 630 & 1200 & 2100\\
R (5y) & 225 & 570 & 940 & 1500\\
R (7y) & 235 & 570 & 850 & 1250\\
R (10y) & 235 & 570 & 850 & 1100\\
$\rec$ & 40 & 40 & 25 & 15\\
S & 2.898 & 2.297 & 2.237 & 2.237\\
$\sigma$ (\%) & 5 & 14 & 20 & 50\\
\hline
\end{tabular}
\end{center}

\caption{\small Input data. }\label{PARhistory}
\end{table}

Some important remarks on the data. First of all we see that the
CDS rates $R$ (which were already high in the first dates we
considered) grow very fast and the recovery rate decreases as the
crisis approaches. The historical volatility we use as a proxy for
1y volatility increases a lot in time as well (probably the last
value of 50\% is even underestimated, considering that after a
couple of days the price has gone below one euro). We have also
set the payout ratio identically equal to zero.

\begin{figure}[!h]
\begin{center}
\includegraphics[angle=270,width=0.8\textwidth]{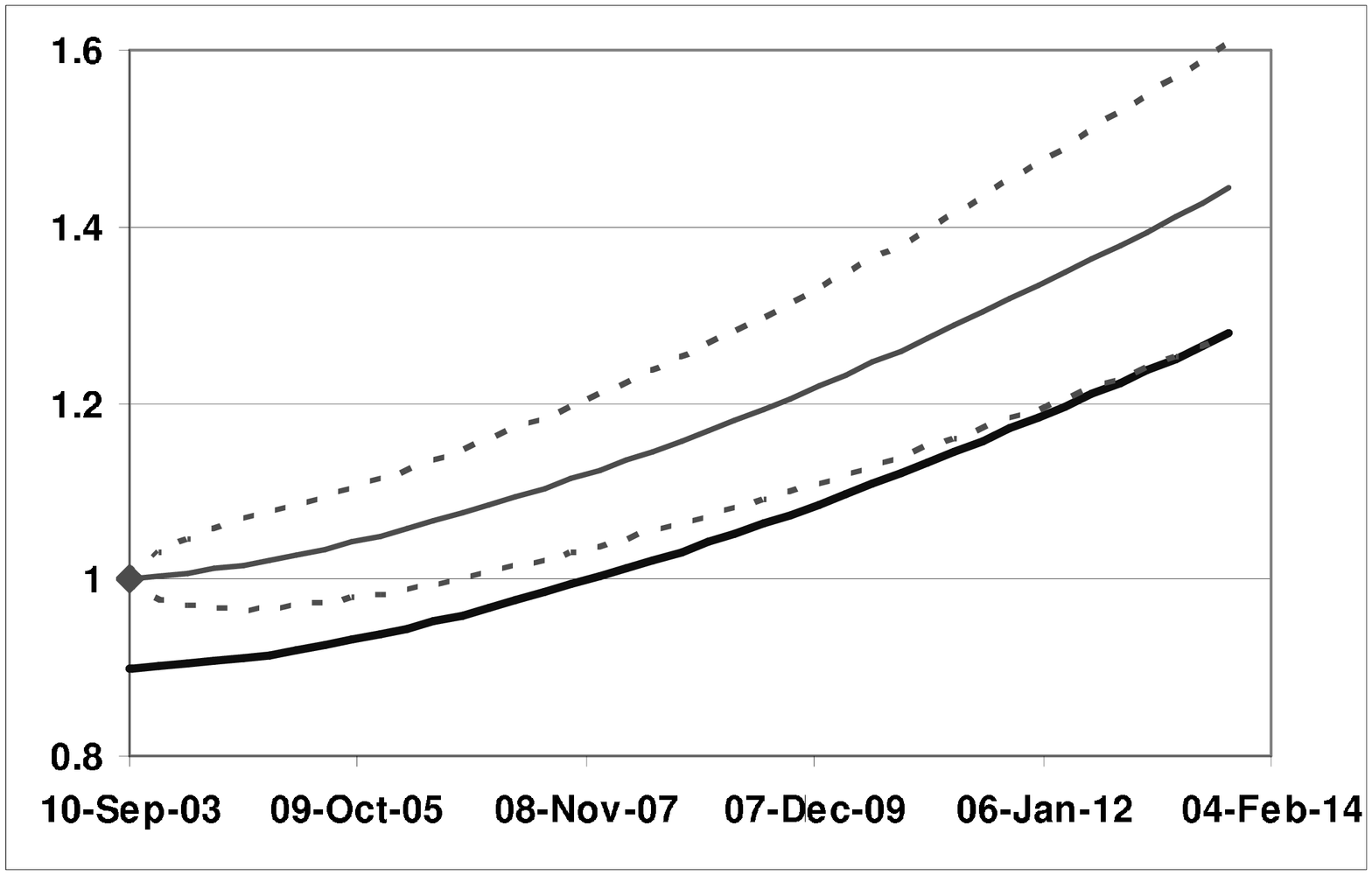}
\end{center}
\caption{\small Default barrier $t\mapsto \Hhat(t)$ on September
10th, 2003, $t\mapsto \Ex(V_t)$ and $t\mapsto
\exp\left(\mathbb{E}[\ln(V_t)]\mp
\mbox{Std}[\ln(V_t)]\right)$.}\label{barrier10sep}
\end{figure}

\begin{figure}[!h]
\begin{center}
\includegraphics[angle=270,width=0.8\textwidth]{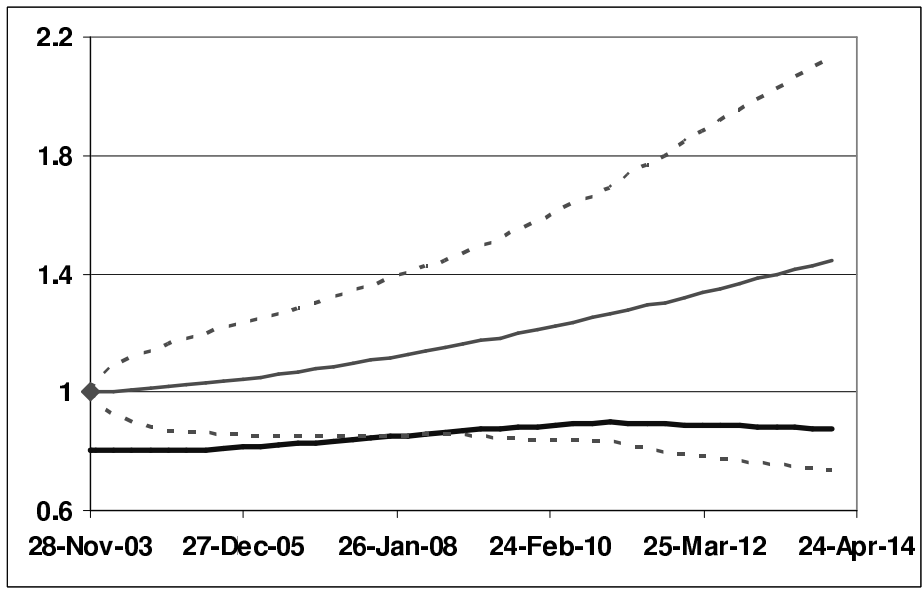}
\end{center}
\caption{\small Same as above on November 28th, 2003.
}\label{barrier28nov}
\end{figure}

\begin{figure}[!h]
\begin{center}
\includegraphics[angle=270,width=0.8\textwidth]{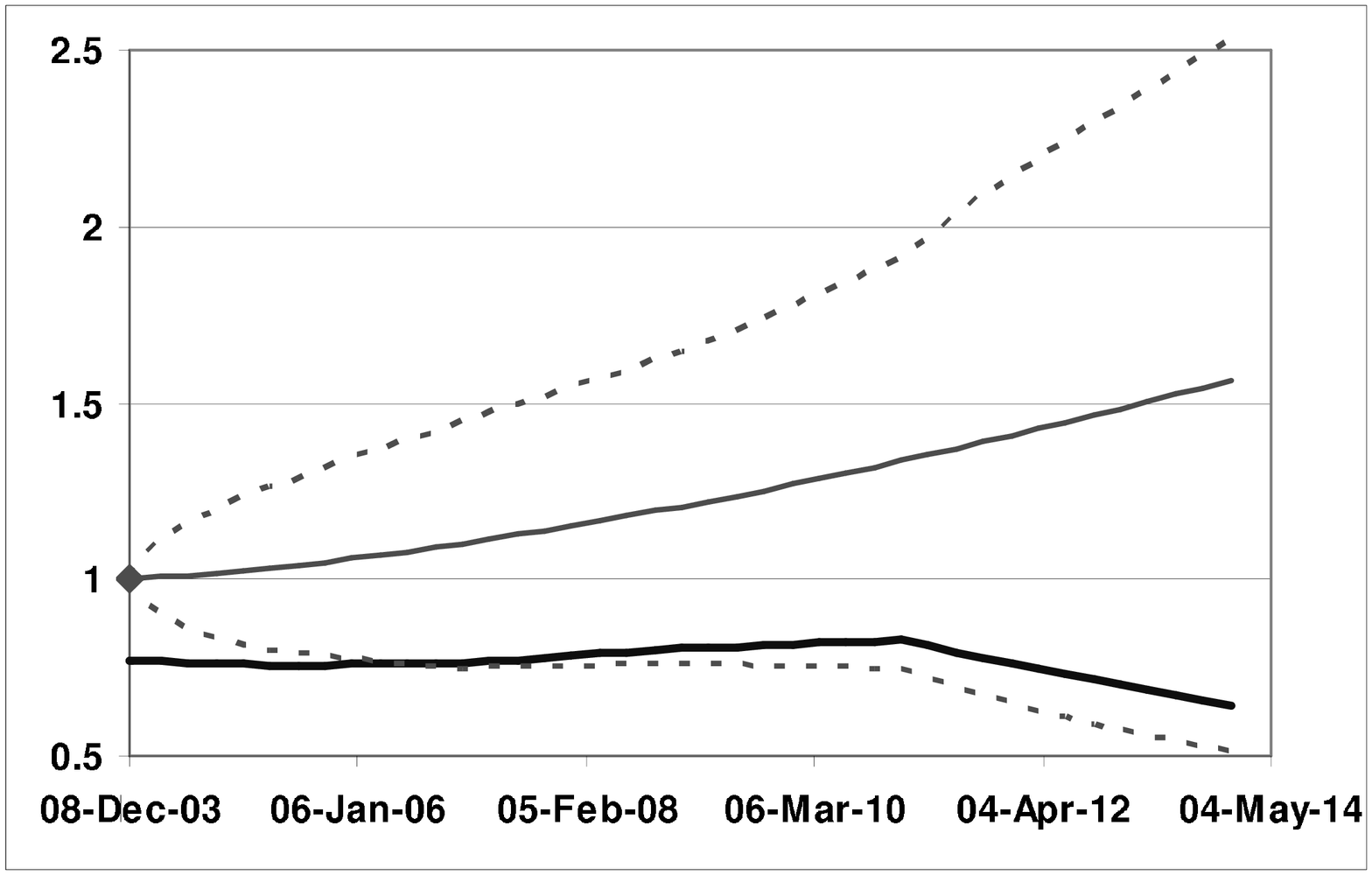}
\end{center}
\caption{\small Same as above on December 8th, 2003.
}\label{barrier8dec}
\end{figure}

\begin{figure}[!h]
\begin{center}
\includegraphics[angle=270,width=0.8\textwidth]{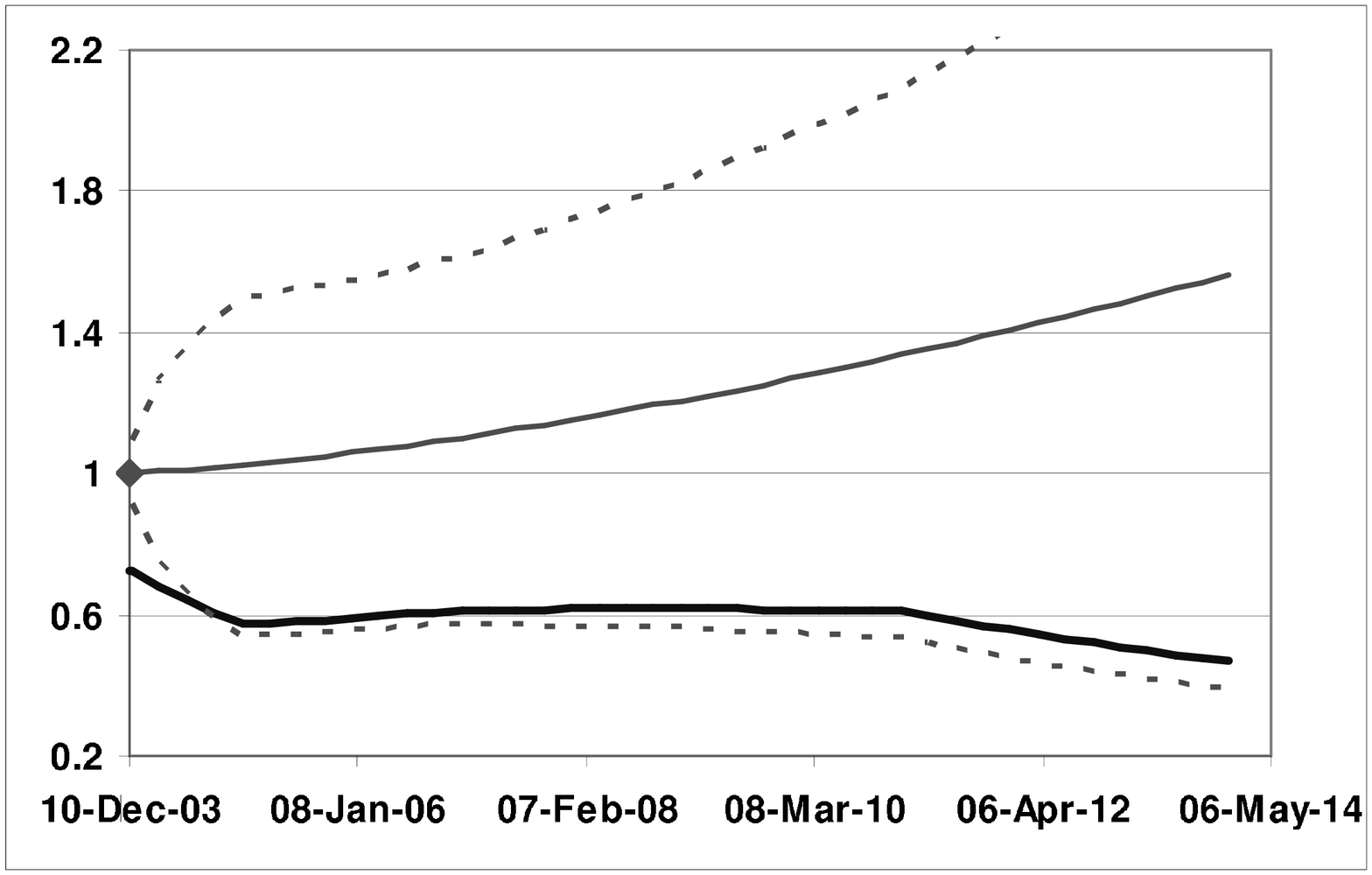}
\end{center}
\caption{\small Same as above on December 10th,
2003.}\label{barrier10dec}
\end{figure}

With this data, we are now able to compute the credit spread based
barrier parameter $H$ for each considered date.
%
%
In principle we could plot the value $H$ on a graph with the
corresponding initial value $V_0$ (either normalized or not), and
also the behavior of the real barrier $\Hhat$.
Even if we did this for each date, this would not be sufficient to
deduce whether the model efficiently incorporates default
information because the different volatilities may be misleading.
More precisely, for normalized values about $1$ we find $H=0.8977$
for September 10th and $H=0.7253$ for December 10th. As the first
safety barrier is closer to the initial value $V_0 = 1$, one could
think that the default is more likely in the first situation, but
we know that this is not the case. The fact is that in the first
situation we have a much lower volatility than in the second. So
we decided to plot not only the barrier $\Hhat$ and the initial
value of the firm $V_0$, but also the expected price in the future
$t \mapsto \Ex(V_t)$ and a confidence region $t \mapsto
\exp(\Ex[\ln(V_t)]- \mbox{Std}[\ln(V_t)]  ) $ and $t \mapsto
\exp(\Ex[\ln(V_t)]+ \mbox{Std}[\ln(V_t)]  )$, where Std is the
risk neutral standard deviation. Std is easily computed for normal
variates such as $\ln(V)$. All these graphs are reported in
Figures from \ref{barrier10sep} to \ref{barrier10dec}. In the
confidence region between the expected price minus/plus one
standard deviation there are the 68\% of the total scenarios.
Hence the instant in which the lower bound of the confidence
region hits the barrier is a signal of the proximity of the
default. And from the graphs it is clear that as time passes, the
default becomes more and more likely and the scenarios
trajectories widen, hitting the barrier earlier. Indeed, if on
September 10th the confidence region hits the barrier after nearly
ten years, on December 10th this happens after less than one year.
At this point of the discussion a few remarks on the choice of
$\beta$ are in order.

Putting the consistency condition (\ref{coherpunct}) in the
equation for the barrier $\Hhat$ given in~(\ref{realbarrier}),
after some calculations we find
\begin{equation}\label{explicitbarrier}
\Hhat(t)=\frac{H}{V_0}\,\mathbb{E}[V_t]\,\exp\left[-\frac{(1+2\beta)}{2}\int_0^t\sigma_s^2
ds\right].
\end{equation}

\begin{remark}\label{remark-beta-par} {\bf The role of the $\beta$ parameter.}
The integral of the squared volatility has the effect of
introducing a concave and decreasing factor into the shape of the
safety barrier, provided the round brackets term is positive.  The
role of $\beta$ is thus clear: For given values of the
volatilities $\sigma$, $\beta$ is the parameter that when
increased allows us to increase the concavity and the steepness of
the decreasing factor in the barrier, thus lowering it.  This
parameter may be useful when the calibrated volatilities get out
of control. Indeed, in case the needed concavity is high and
$\beta$ is small, volatilities may tend to explode to obtain the
required barrier factor. By increasing $\beta$ the volatilities
needed to produce a given steepness and concavity can become
smaller, avoiding this risk.
\end{remark}

In our first experiments we choose $\beta = 0.5$ such that the
argument of the exponential is simply the integral of the
variance, but this is just our particular choice based on
numerical experiments. In this case, for low values of the
volatilities the barrier is very similar to the behavior of the
expected price, hence it is upward sloped. On the contrary, high
values of the volatilities change significantly the shape of the
barrier, but also the confidence region widens as a consequence.
These effects can be checked by comparing Figures from
\ref{barrier10sep} to \ref{barrier10dec} to the Tables reported in
the Appendix. In Figure \ref{volats8dec} we show a typical term
structure of volatility obtained from the calibration,
corresponding to the data of December 8th.

\begin{figure}[!h]
\begin{center}
\includegraphics[angle=270,width=0.8\textwidth]{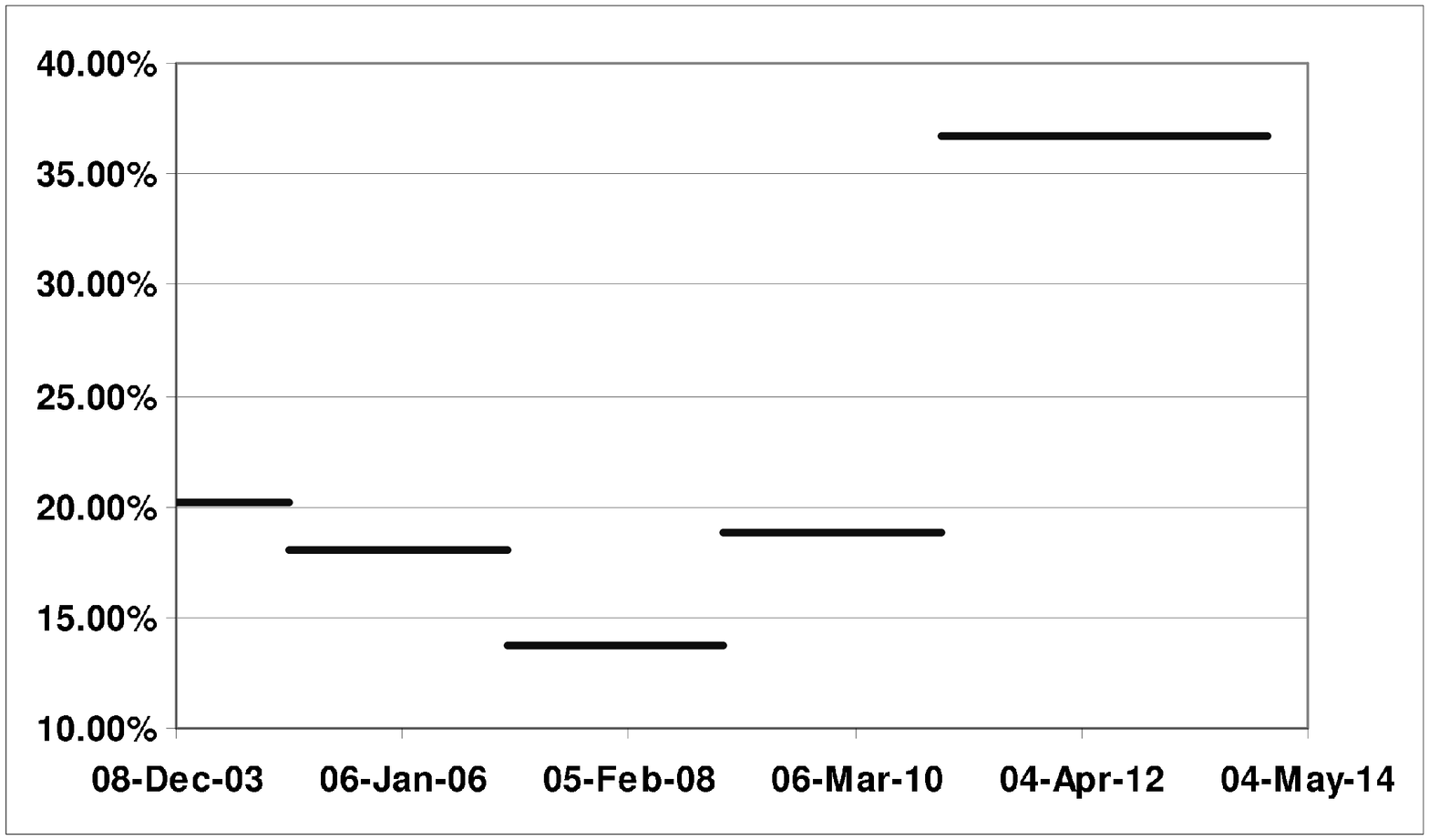}
\end{center}
\caption{\small Volatility term structure $T \mapsto \sigma(T)$ on
December 8th, 2003. }\label{volats8dec}
\end{figure}

\begin{figure}[!h]
\begin{center}
\includegraphics[angle=270,width=0.8\textwidth]{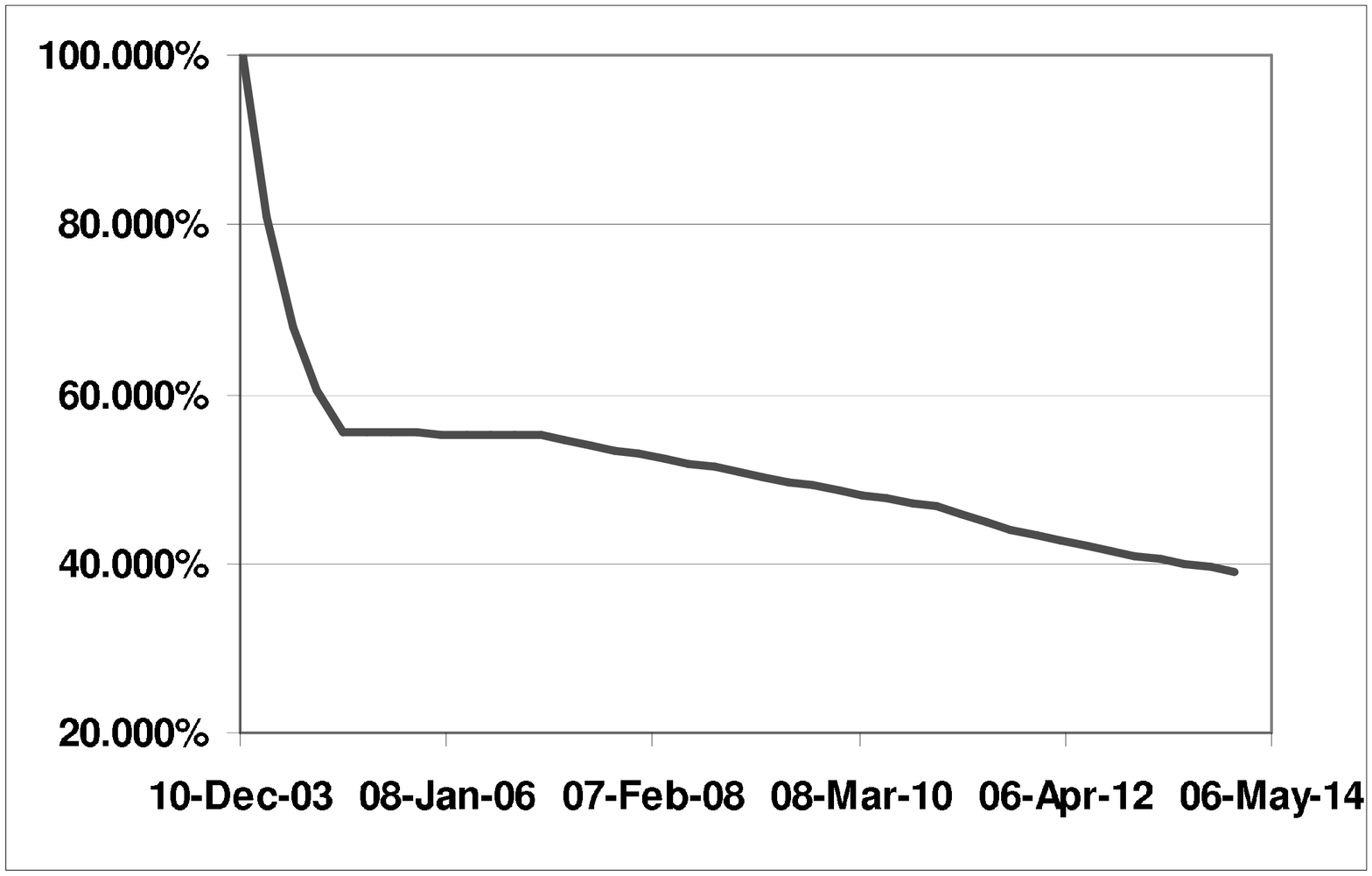}
\end{center}
\caption{\small Survival probability on December 10th,
2003.}\label{surv}
\end{figure}

Furthermore, in Figure \ref{surv} we plot the survival probability
on December 10th. We see that the probability falls down very
quickly in the first period, but later it decreases in a much
slower way. This fact can be explained as follows. When the firm
is approaching default, investors are more interested in buying
protection over a small time horizon, because they assess a very
high probability to the default event for the immediate future
period. But if the firm survives to the full crisis, probably it
will reach a certain stability situation, characterized by a
``lower" default probability, so that provided the company
survives the first immediate times after the crisis, investors are
less worried by subsequent periods.

As a final investigation, we have repeated the calibration when
$H$ is selected under the alternative method of the
``excursion/protection analogy", i.e. by setting $H/V_0 = \rec$.
In the Appendix we report all the results. The first important
remark is that here we have used $\beta=0.08$. This is due to the
fact that the calibration for the last date requires very high
volatilities, being the relevant $H$ very low. Then we are
compelled to use low $\beta$'s to avoid the calibration explosion
hinted at in Remark~\ref{remark-beta-par}. At the same time we
cannot diminish $\beta$ too much, because the second volatility
$\sigma(2y\div3y)$ is very low due to the peculiar structure of
CDS market quotes across maturity. Ceteris paribus, diminishing
$\beta$ means raising the barrier and thus diminishing
volatilities, but $\sigma$ has a lower bound equal to zero. So, in
the end, the feasible interval for the $\beta$ values is rather
narrow and $\beta=0.08$ is one value in this narrow interval. For
previous dates, the calibration was possible also with other
values of $\beta$ but we decided to keep it fixed to $0.08$ for
uniformity reasons.

From the results (Tables \ref{output10sep_bis} to
\ref{output10dec_bis}), we see that in this second framework, the
stripped volatilities are much higher than before, especially in
proximity of the crisis, where they are suspiciously high for
practical purposes and raise a number of questions on realism.
However, this effect is essentially due to the low values of $H$
and not to $\beta$, as one could suspect. Indeed, in Table
\ref{output10sep_ter} we present the results of a calibration
performed for the first date, with $\beta=0.08$ and $H$ computed
with the ``credit spread based" method, and it is immediately
clear that the volatilities are no longer pathologically large and
are nearly the same as in Table \ref{output10sep}. Anyway, the
survival probabilities are not sensibly influenced by the method
chosen to estimate $H$ when the CDS to be calibrated are fixed;
this lack of sensitivity is mainly due to the peculiar role of
$\beta$ and to the possibility to adjust its value, which in
practice is an important degree of freedom of the model.

\begin{remark}{\bf (SBAT1P, Scenario Barrier version of the AT1P model)}
A final remark on the possible use of random scenario-based
barriers is in order. It is well known, see for example Bielecki
and Rutkowski (2001), that in structural models with deterministic
threshold barriers the default time is predictable. This is at
times considered a drawback, especially in multi-name situations
where one has to take into account contagion effects (Giesecke
(2002)). What can help in this case is a random scenario-based
default barrier. We may see that in our specific case the
introduction of a random variable for the $H$ parameter,
independent of the driving Brownian Motion $W$, induces through
iterated conditioning an analytical formula for default
probabilities and thus CDS's. This formula is based on a one
dimensional integration of (\ref{survcoher}) in $H$ against the
density of the random variable replacing the deterministic $H$. We
call this model the SBAT1P (Scenario Barrier Analytically
Tractable 1st Passage model) and we investigate its use in Brigo
and Tarenghi (2005), where we use parameters in the distribution
of $H$ to try and calibrate CDS prices. Scenarios on value of the
firm volatilities can be employed as well (SVBAT1P model).
\end{remark}

\section{A fundamental example: Pricing Counterparty Risk in Equity Swaps}\label{sec:otherproducts}
In this section we present an example of pricing with the
calibrated structural model. This example concerns the valuation
of an equity swap where we take into account counterparty risk,
and is chosen to highlight one case where the calibrated
structural model may be preferable to a reduced-form intensity
model calibrated to the same market information. This is an
illustration of a more general situation that typically occurs
when one tries to price the counterparty risk in an equity payoff.
We will see that it is possible to split the expectation of the
payoff, and that the decomposition roughly involves the valuation
of the same payoff without counterparty risk and the valuation of
an option on the residual NPV of the considered payoff at the
default time of the counterparty. Therefore including the
counterparty risk adds an optionality level to the payoff.

Let us consider an equity swap payoff. Assume we are a company
``A" entering a contract with company ``B", our counterparty. The
reference underlying equity is company ``C". The contract, in its
prototypical form, is built as follows. Companies ``A" and ``B"
agree on a certain amount $K$ of stocks of a reference entity ``C"
(with price $S$) to be taken as nominal ($N=K\,S_0$). The contract
starts in $T_a=0$ and has final maturity $T_b = T$. At $t=0$ there
is no exchange of cash (alternatively, we can think that ``B"
delivers to ``A" an amount $K$ of ``C" stock and receives a cash
amount equal to $K S_0$). At intermediate times ``A" pays to ``B"
the dividend flows of the stocks (if any) in exchange for a
periodic rate (for example a semi-annual LIBOR or EURIBOR rate
$L$) plus a spread $X$. At final maturity $T=T_b$, ``A" pays $K
S_T$ to ``B" (or gives back the amount $K$ of stocks) and receives
a payment $K S_0$. This can be summarized as follows:

\begin{center}
Initial Time 0: no flows, or\\  A $\longrightarrow$ $K S_0$ cash
$\longrightarrow$ B\\
 A $ \longleftarrow$  $K$ equity $\longleftarrow$ B\\
....  \\
Time $T_i$:\\ A $\longrightarrow$ equity dividends $\longrightarrow$ B\\
A $\longleftarrow$  Libor + Spread $\longleftarrow$ B\\
.... \\
Final Time $T_b$:\\  A $\longrightarrow$ K equity
$\longrightarrow$ B
\\
A $\longleftarrow$  $K S_0$ cash $\longleftarrow$ B \end{center}

The price of this product can be derived using risk neutral
valuation, and the (fair) spread is chosen in order to obtain a
contract whose value at inception is zero. We ignore default of
the underlying ``C", thus assuming it has a much stronger credit
quality than the counterparty ``B". This can be the case for
example when ``C" is an equity index (Pignatelli~(2004)). It can
be proved that if we do not consider default risk for ``B", the
fair spread is identically equal to zero. But when taking into
account counterparty default risk in the valuation the fair spread
is no longer zero.   In case an early default of the counterparty
``B" occurs, the following happens.  Let us call $\tau=\tau_B$ the
default instant. Before $\tau$ everything is as before,  but if
$\tau\leq T$, the net present value (NPV) of the position at time
$\tau$ is computed. If this NPV is negative for us, i.e. for ``A",
then its opposite is completely paid to ``B" by us at time $\tau$
itself. On the contrary, if it is positive for ``A", it is not
received completely but only a recovery fraction $\rec$ of that
NPV is received by us. It is clear that to us (``A") the
counterparty risk is a problem when the NPV is large and positive,
since in case ``B" defaults we receive only a fraction of it.

Analytically, the risk neutral expectation of the discounted
payoff is ($L(S,T)$ is the simply compounded rate at time $S$ for
maturity $T$):
\begin{eqnarray}\label{ESpayoff}
\Pi_{ES}(0) & = &
\mathbb{E}_0\bigg\{\mathbf{1}_{\{\tau>T_b\}}\bigg[-K\,\npv_{dividends}^{0\div
T_b}(0)+ KS_0\,\sum_{i=1}^b D(0,T_i)\alpha_i\big(L(T_{i-1},T_i)
\nonumber\\
& & +X\big)+D(0,T_b)\big(KS_0-KS_{T_b}\big)\bigg] \nonumber\\
& &  +\mathbf{1}_{\{\tau\leq
T_b\}}\bigg[-K\,\npv_{dividends}^{0\div
\tau}(0)+KS_0\,\sum_{i=1}^{\beta(\tau)-1}D(0,T_i)\alpha_i\big(L(T_{i-1},T_i)
\nonumber\\
& & +X\big)+D(0,\tau)\big(\rec(\npv(\tau))^+
-(-\npv(\tau))^+\big)\bigg]\bigg\}
\end{eqnarray}
where
\begin{eqnarray}\label{NPVdef}
\npv(\tau) & = &
\mathbb{E}_{\tau}\bigg\{-K\,\npv_{dividends}^{\tau\div
T_b}(\tau)+KS_0\,\sum_{i=\beta(\tau)}^{b}
D(\tau,T_i)\alpha_i\left(L(T_{i-1},T_i)+X\right) \nonumber \\
& & +\left(KS_0-KS_{T_b}\right)D\left(\tau,T_b\right)\bigg\}.
\end{eqnarray}
We denote by $\npv_{dividends}^{s \div t}(u)$  the net present
value of the dividend flows between $s$ and $t$ computed in $u$.

In the following we will prove the
\begin{proposition}{\bf (Equity Return Swap price under
Counterparty Risk)}. The fair price of the Equity Swap defined
above, i.e.~(\ref{ESpayoff}), can be simplified as follows:
\begin{eqnarray*}
\Pi_{ES}(0) = KS_0 X \sum_{i=1}^b\alpha_i
P(0,T_i)-\lgd\,\mathbb{E}_0\bigg\{ \mathbf{1}_{\{\tau\leq
T_b\}}P(0,\tau)(\npv(\tau))^+\bigg\}.
\end{eqnarray*}
The first term is the equity swap price in a default-free world,
whereas the second one is the optional price component due to
counterparty risk.
\end{proposition}

\begin{remark}\label{remarkaccruin} {\bf (Different choice for the NPV)}.
Let us focus for a second on the NPV term in (\ref{NPVdef}). In
particular let us concentrate on the interval
$[T_{\beta(\tau)-1},T_{\beta(\tau)})$ containing the default time
$\tau$. If $\tau$ is strictly inside the interval, the LIBOR +
spread resetting at the earlier $T_{\beta(\tau)-1}$ is paid for
the whole interval.

A different formulation is the following. In case the default time
is inside a certain interval, the LIBOR + spread term for the part
of this interval preceding default is immediately paid and we add
the remaining part, to be paid at the end of the interval,
discounted back from the end of the interval to the current
(default) time. This amounts to let the summation in
(\ref{NPVdef}) start from $\beta(\tau)+1$ rather than from
$\beta(\tau)$ and to add the following term in the NPV payoff:
\[\left(L\left(T_{\beta(\tau)-1},T_{\beta(\tau)}\right)+X\right)\left(\tau-T_{\beta(\tau)-1}\right)
+
\left(L\left(T_{\beta(\tau)-1},T_{\beta(\tau)}\right)+X\right)\left(T_{\beta(\tau)}-\tau\right)D\left(\tau,T_{\beta(\tau)}\right)
\]
Formulation (\ref{NPVdef}) reads like postponing the default
$\tau$ to the first $T_i$ following $\tau$ and then discounting it
back to the current (default) time.

Both formulations make sense so contract specifications are
important in order to price this kind of products. Analytical
pricing formulas with our choice result to be more tractable than
with the alternative choice presented in this remark, where small
accruing terms appear. Such accruing terms are very small in
general and can be ignored.
\end{remark}

If we try and find the above price by computing the expectation
through a Monte Carlo simulation, we have to simulate both the
behavior of the equity ``C" underlying the swap, which we call
$S_t=S_t^C$,  and the default of the counterparty ``B". In
particular we need to know exactly $\tau=\tau_B$. Obviously the
correlation between ``B" and ``C" could have a relevant impact on
the contract value. Here the structural model can be helpful:
Suppose to calibrate the underlying process $V$ to CDS's for name
``B", finding the appropriate default barrier and volatilities
according to the procedure outlined earlier in this paper with the
AT1P model. We could set a correlation between the processes
$V^B_t$ for ``B" and $S_t$ for ``C", derived for example through
historical estimation directly based on equity returns, and
simulate the joint evolution of $[V^B_t, S_t]$. As a proxy of the
correlation between these two quantities we may consider the
correlation between $S^B_t$ and $S^C_t$, i.e. between equities.

\begin{remark}{\bf (Using reduced form models?)}
At this point it may be interesting to consider the alternatives
offered by an intensity reduced form model. If one takes a
deterministic intensity, since Poisson processes and Brownian
Motions defined on the same space are independent, there is no way
to introduce a correlation between the default event for ``B" and
the value of the equity ``C". To take into account this
correlation we may have to move to stochastic intensity. However,
correlating the stochastic intensity of ``B" to the equity of ``C"
may lead to a poorly tractable model, even under tractable
intensities as in Brigo and Alfonsi (2005), and the amount of
correlation induced in this way may be too small for practical
purposes. Furthermore, historical estimation of the correlation
between the instantaneous intensity and the equity of ``C" may
pose some problems that the structural model avoids by definition.
\end{remark}

Now let us go more into details in analyzing the payoff. For
simplicity, we assume deterministic dividends and rates, so that
we can substitute the discount factors $D(t,T)$ with zero coupon
bond prices $P(t,T)$ and move them outside the expectations.
Moreover we recall that $L(T_{i-1},T_i) =
\left(P(0,T_{i-1})/P(0,T_i)-1\right)/\alpha_i$ and, as a
consequence, $\sum_{\small i=A+1}^{\small B} P(0,T_i)\alpha_i
L(T_{i-1},T_i)=P(0,T_A) - P(0,T_B)$. Finally, we set $\rec =
1-\lgd$, consistently with CDS notation, and notice that
\begin{equation}\label{npvpospart}
\rec\,(\npv(\tau))^+ -(-\npv(\tau))^+ = \npv(\tau)
-\lgd(\npv(\tau))^+.
\end{equation}
We can rewrite the discounted NPV as
\begin{eqnarray}\label{NPVsimpl}
P(0,\tau) \npv(\tau) & = & -K\,\npv_{dividends}^{\tau\div
T_b}(0)+KS_0\,\sum_{i=\beta(\tau)}^{b}
P(0,T_i)\alpha_i(L(T_{i-1},T_i)+X) \nonumber \\
& &+KS_0 P(0,T_b) - K\mathbb{E}_{\tau}\{S_{T_b}\}P(0,T_b).
\end{eqnarray}
Substituting (\ref{npvpospart}) and (\ref{NPVsimpl}) in
(\ref{ESpayoff}) we obtain:
\begin{eqnarray}\label{ESpayoffsimpl}
\nonumber \Pi_{ES}(0) & = &
\mathbb{E}_0\bigg\{\mathbf{1}_{\{\tau>T_b\}}\bigg[-K\,\npv_{dividends}^{0\div
T_b}(0)+ K S_0 \sum_{i=1}^b P(0,T_i) \alpha_i (L(T_{i-1},T_i)+X)\\
\nonumber & & + P(0,T_b)(KS_0-KS_{T_b})\bigg]\\
\nonumber \ \ \ & & + \mathbf{1}_{\{\tau\leq
T_b\}}\bigg[-K\,\npv_{dividends}^{0\div
\tau}(0)+KS_0\sum_{i=1}^{\beta(\tau)-1} P(0,T_i) \alpha_i (L(T_{i-1},T_i)+X)\\
\nonumber & &+
P(0,\tau)\npv(\tau)-\lgd\,P(0,\tau)(\npv(\tau))^+\bigg]\bigg\} =\\
\nonumber & = &
\mathbb{E}_0\bigg\{\mathbf{1}_{\{\tau>T_b\}}\bigg[-K\,\npv_{dividends}^{0\div
T_b}(0)+ K S_0 \sum_{i=1}^b P(0,T_i) \alpha_i (L(T_{i-1},T_i)+X)\\
\nonumber & & + P(0,T_b)KS_0-P(0,T_b)KS_{T_b}\bigg]\\
\nonumber & & + \mathbf{1}_{\{\tau\leq
T_b\}}\bigg[-K\,\npv_{dividends}^{0\div
T_b}(0)+KS_0\sum_{i=1}^b P(0,T_i) \alpha_i (L(T_{i-1},T_i)+X)\\
\nonumber & & + P(0,T_b) KS_0 - P(0,T_b)
K\mathbb{E}_{\tau}\big\{S_{T_b}\big\}-\lgd\,P(0,\tau)(\npv(\tau))^+\bigg]\bigg\}=\\
\nonumber & = & -K\,\npv_{dividends}^{0\div T_b}(0)+KS_0\sum_{i=1}^b P(0,T_i) \alpha_i (L(T_{i-1},T_i)+X)\\
\nonumber & &
+P(0,T_b)KS_0-P(0,T_b)K\mathbb{E}_0\bigg\{\mathbf{1}_{\{\tau>T_b\}}S_{T_b}\bigg\}
- P(0,T_b) K \mathbb{E}_0\bigg\{\mathbf{1}_{\{\tau\leq
T_b\}}\mathbb{E}_{\tau}\{S_{T_b}\}\bigg\}\\
\nonumber & & -\lgd\,\mathbb{E}_0\bigg\{ \mathbf{1}_{\{\tau\leq
T_b\}}P(0,\tau)(\npv(\tau))^+\bigg\} =\\ \nonumber & = &
-K\,\npv_{dividends}^{0\div T_b}(0)+ KS_0 +KS_0 X
\sum_{i=1}^b\alpha_i P(0,T_i)\\ \nonumber & &
-P(0,T_b)K\mathbb{E}_0\bigg\{\mathbf{1}_{\{\tau>T_b\}}S_{T_b}+\mathbf{1}_{\{\tau\leq
T_b\}}\mathbb{E}_{\tau}\{S_{T_b}\}\bigg\}\\
& & -\lgd\,\mathbb{E}_0\bigg\{ \mathbf{1}_{\{\tau\leq
T_b\}}P(0,\tau)(\npv(\tau))^+\bigg\}.
\end{eqnarray}
Now, observing that
\begin{eqnarray}\label{changeindicator}
\mathbb{E}_0\big\{\mathbf{1}_{\{\tau>T_b\}}S_{T_b}+\mathbf{1}_{\{\tau\leq
T_b\}}\mathbb{E}_{\tau}\{S_{T_b}\}\big\} & = &
\mathbb{E}_0\big\{\mathbf{1}_{\{\tau>T_b\}}S_{T_b}+(1-\mathbf{1}_{\{\tau>
T_b\}})\mathbb{E}_{\tau}\{S_{T_b}\}\big\} = \nonumber\\
 & = & \mathbb{E}_0\big\{\mathbb{E}_{\tau}\{S_{T_b}\}+\mathbf{1}_{\{\tau>T_b\}}\big(S_{T_b}-
\mathbb{E}_{\tau}\{S_{T_b}\}\big)\big\}= \nonumber \\ & = &
\mathbb{E}_0\big\{\mathbb{E}_{\tau}\{S_{T_b}\}\big\}
+\mathbb{E}_0\big\{\mathbf{1}_{\{\tau>T_b\}}(S_{T_b}-S_{T_b})\big\}=
\nonumber\\ & = &\mathbb{E}_0\big\{S_{T_b}\big\} =
\frac{S_0}{P(0,T_b)}-\frac{\npv_{dividends}^{0\div
T_b}(0)}{P(0,T_b)}
\end{eqnarray}
the expected payoff (\ref{ESpayoffsimpl}) becomes
\begin{eqnarray}\label{plainES} \Pi_{ES}(0) & = &
-K\,\npv_{dividends}^{0\div T_b}(0)+ KS_0 +KS_0 X
\sum_{i=1}^b\alpha_i P(0,T_i)\nonumber\\
& &
-K\,P(0,T_b)\left(\frac{S_0}{P(0,T_b)}-\frac{\npv_{dividends}^{0\div
T_b}(0)}{P(0,T_b)}\right)\nonumber\\
& & -\lgd\,\mathbb{E}_0\bigg\{ \mathbf{1}_{\{\tau\leq
T\}}P(0,\tau)(\npv(\tau))^+\bigg\} = \nonumber\\
& = & KS_0 X \sum_{i=1}^b\alpha_i
P(0,T_i)-\lgd\,\mathbb{E}_0\bigg\{ \mathbf{1}_{\{\tau\leq
T_b\}}P(0,\tau)(\npv(\tau))^+\bigg\}.
\end{eqnarray}
which proves our proposition above. As noticed in the proposition,
the first term in the last expression in (\ref{plainES}) is
$\Pi_{ES}^{\defree}(0)$, i.e. the payoff of the Equity Swap in a
default free (``defree") world. Indeed, as stated before, in a
default free market the fair spread would be zero, so that if we
add a nonzero spread $X$ the value of the equity swap in a default
free world is simply the NPV of the future spread cash flows.

The second term is very important if we allow counterparty ``B" to
default. As it is non-positive, to have a fair value for
$\Pi_{ES}(0)$ we need a non-negative spread $X$.

At this point it may be worth noticing that the above
decomposition is rather general. When we include counterparty risk
into the valuation, we always obtain a decomposition
\begin{center}
{\small Price = Price{\tiny DefaultFree} - (1-recovery)
CallOption{\tiny on residual NPV}}
\end{center}
The call option has strike 0. As we have observed earlier, the
counterparty risk introduces, among other features, a further
optionality level into the valuation.

Going back to our equity swap, now it is possible to run the Monte
Carlo simulation, looking for the spread $X$ that makes the
contract fair. The simulation itself is simpler when taking into
account the following computation included in the discounted NPV:
\begin{equation}\label{exptaufinalprice}
P(\tau,T_b)\mathbb{E}_{\tau}\big\{S_{T_b}\big\} = S_{\tau}
-\npv_{dividends}^{\tau\div T_b}(\tau)
\end{equation}
so that we have
\begin{eqnarray}\label{NPVsimpl2}
\nonumber P(0,\tau) \npv(\tau) & = &
KS_0\,\sum_{i=\beta(\tau)}^{b} P(0,T_i)\alpha_i(L(T_{i-1},T_i)+X)+KS_0 P(0,T_b)\\
& & \nonumber - K P(0,\tau) S_{\tau} =\\
\nonumber & = & KS_0\,\sum_{i=\beta(\tau)}^{b} P(0,T_i)\alpha_i X
+ KS_0 P(0,T_{\beta(\tau)-1})\\ & &  - K P(0,\tau) S_{\tau}.
\end{eqnarray}

The reformulation of the original expected payoff (\ref{ESpayoff})
as in (\ref{plainES}) presents an important advantage in terms of
numerical simulation. In fact in (\ref{ESpayoff}) we have a global
expectation, hence we have to simulate the exact payoff for each
path. In (\ref{plainES}), with many simplifications, we have
isolated a part of the expected payoff out of the main
expectation. This isolated part has an expected value that we have
been able to calculate, so that it does not have to be simulated.
Simulating only the residual part is helpful because now the
variance of the part of the payoff that has been computed
analytically is no longer affecting the standard error of our
Monte Carlo simulation. The standard error is indeed much lower
when simulating (\ref{plainES}) instead of (\ref{ESpayoff}). The
expected value we computed analytically above involves terms in
$S_{T_b}$ which would add a lot of variance to the final payoff.
In (\ref{plainES}) the only $S_{T_b}$ term left is in the optional
NPV part.

We performed some simulations under different assumptions on the
correlation between ``B" and ``C". We considered five cases:
$\rho= -1$, $\rho = -0.2$, $\rho = 0$, $\rho = 0.5$ and $\rho =
1$. In Table \ref{EScorrelation} we present the results of the
simulation, together with the error given by one standard
deviation (Monte Carlo standard error). For counterparty ``B" we
used the same CDS rates of the company analyzed in Section
\ref{sec:calibration}. For the reference stock ``C" we used a
hypothetical stock with initial price $S_0=20$, volatility $\sigma
= 20\%$ and constant dividend yield $q=0.80\%$. The contract has
maturity $T=5y$ and the settlement of the LIBOR rate has a
semi-annual frequency. Finally, we included a recovery rate
$\rec=40\%$. The starting date is the same we used for the
calibration, i.e. March 10th, 2004. Since the reference number of
stocks $K$ is just a constant multiplying the whole payoff,
without losing generality we set it equal to one.

In order to reduce the errors of the simulations, we have adopted
a variance reduction technique using the default indicator (whose
expected value is the known default probability) as a control
variate. In particular we have used the default indicator
$1_{\{\tau < T\}}$ at the maturity $T$ of the contract, which has
a large correlation with the final payoff. Even so, a large number
of scenarios are needed to obtain errors with a lower order of
magnitude than $X$. In our simulations we have used $N = 2000000$.

We notice that $X$ increases together with $\rho$. This fact can
be explained in the following way. Let us consider the case of
positive correlation between ``B" and ``C": This means that, in
general, if the firm value for ``B" increases, moving away from
the default barrier, also the stock price for ``C" tends to
increase due to the positive correlation. Both processes will then
have high values. Instead, again under positive correlation, if
$V^B_t$ lowers towards the default barrier, also $S^C_t$ will tend
to do so, going possibly below the initial value $S_0$. In this
case $NPV(\tau)$ has a large probability to be positive (see
(\ref{NPVsimpl2})), so that one needs a large $X$ to balance it,
as is clear when looking at the final payoff (\ref{plainES}). On
the contrary, for negative correlation, the same reasoning can be
applied, but now if $V^B_t$ lowers and tends to the  default
barrier, in general $S^C_t$ will tend to move in the opposite
direction and the corresponding $NPV(\tau)$ will probably be
negative, or, if positive, not very large. Hence the ``balancing"
spread $X$ we need will be quite small.

\begin{table}[h!]
\begin{center}
\begin{tabular}{|c|c|c|c|}
\hline $\rho$ & X & ES payoff & MC error \\
\hline
-1      &   0       &   0       &   $0^*$   \\
-0.2    &   2.45    &   -0.02   &   1.71    \\
0       &   4.87    &   -0.90   &   2.32    \\
0.5     &   14.2    &   -0.53   &   2.71    \\
1       &   24.4    &   -0.34   &   0.72    \\
\hline
\end{tabular}
\end{center}
\caption{\small Fair spread $X$ (in basis points) of the Equity
Swap in five different correlation cases, $S_0=20$. We also report
the value of the average of the simulated payoff (times 10000)
across the $2000000$ scenarios and its standard error, for the
chosen $X$, thus showing that $X$ is indeed fair in practice,
since all the averages and MC windows are close to zero. $X$'s
have been found by iterating the MC simulation for different
values of $X$ until the payoff was found to be sufficiently small.
The MC error for $\rho=-1$ (marked with an asterisk) is null since
in each simulated scenario the simulated term has been found to be
identically zero (i.e. the NPV was always
negative).}\label{EScorrelation}
\end{table}

The simulation error depends on $\rho$ too, and in particular it
is influenced by two main effects. The first one is the same as
before. Indeed, a positive correlation means high probability of
positive NPV in case of default. As the default probability of
``B" does not depend on $\rho$, it is clear that a positive
correlation leads to evaluate the expectation of a random variable
which has a high probability of being different from zero (and in
particular positive), while a negative correlation leads to the
expectation of a variable which is almost always zero, leading to
a lower Monte Carlo error. The other effect is due to the impact
of $\rho$ on the correlation between the payoff variate and the
control variate. It happens that this correlation is decreasing in
$\rho$: It is nearly $-1$ when $\rho=1$, and then increases with
lower values of $\rho$ (it is about $-0.4$ for $\rho = -0.2$).

As a final remark we notice that the spread $X$ corresponding to
$\rho = -1$ is identically equal to zero. The reason is quite
simple: When we have maximum negative correlation between the
counterparty ``B" and the reference entity ``C", the NPV in
(\ref{plainES}) is negative in all the simulated scenarios
involving default, so that the related term in the price of the
equity swap is always zero. We are then left only with the fixed
term and this is null only when $X$ is zero. All this reasoning
also explains why the error of the simulation, marked with an
asterisk, is zero.

\section{Conclusions}
In general the link between default probabilities and credit
spreads is best described by intensity models. The credit spread
to be added to the risk free rate represents a good measure of a
bond credit risk for example. Yet, intensity models present some
drawbacks: They do not link the default event to the economy but
rather to an exogenous jump process whose jump component remains
unexplained from an economic point of view. Moreover, when dealing
with default correlation, a copula function must be introduced
between the jump processes thresholds in a way that has no clear
immediate relation with equity correlation, the only source of
correlation that can be used for practical purposes.

In this paper we introduced an analytically tractable structural
model that allows for a solution to the above points. In this
model the default has an economic cause, in that it is caused by
the value of the firm hitting the safety barrier value, and all
quantities are basic market observables. Also, when dealing with
multi-name products, the model allows for the introduction of the
correlation in a very natural way, by simply correlating shocks in
the different values of the firm equities by means of equity
correlation.


We showed how to calibrate the model parameters to actual market
data: Starting from CDS quotes, we calibrated the value of the
firm volatilities and found the barrier triggering the default
that was consistent with CDS quotes and also leading to analytical
tractability. We also explained the analogies with barrier option
pricing, in particular the case with time dependent parameters. We
hinted at the possible use of a scenario based random barrier, as
in Brigo and Tarenghi (2005).


As a practical example, we also applied the model to a concrete
case, showing how it can describe the proximity of default when
time changes and the market quotes for the CDS's reflect
increasing credit deterioration.  When the market detects a
company crisis, it responds with high CDS quotes and this
translates into high default probabilities, i.e. high
probabilities for the underlying process to hit the safety
barrier, that in turn translate in high calibrated volatilities
for the firm value dynamics. We also took advantage of the
numerical examples to further explain and interpret a few model
parameters whose role needed further investigation.

The calibrated model is useful for pricing more sophisticated
derivatives depending on default. We gave the fundamental example
of equity swaps pricing under counterparty risk.

\section{Appendix: Results of the case study calibrations and market data}
\subsection{Credit spread based barrier parameter $H$}
Here we present the numerical results of the calibration relative
to the case study discussed in Section \ref{sec:casestudy}. For
each date we present the spot price of the equity  with its
historical volatility, the rates and the recovery rates of the
(running) CDS's (quarterly paid), the outputs of the calibration.
For all cases we used $\beta = 0.5$. $H$ is computed using the
credit spread based method explained in Section
\ref{sec:calibration}.
\begin{itemize}
\item September 10th, 2003.\\
Stock Price $= 2.898$ euros,  volatility $=5\%$, recovery rate
$=40\%$. The barrier parameter is $H=0.8977$ (re-scaled $V_0=1$).
\begin{table}[h!]
\begin{center}
\begin{tabular}{|c|c|c|c|}
\hline
CDS maturity & Rate (bps) & Volatility nodes & Survival\\
\hline 10-sep-03 & - & 5.012\% & 100.000\%\\
1y & 192.5 & 5.012\% & 96.673\%\\
3y & 215 & 3.103\% & 89.524\%\\
5y & 225 & 3.178\% & 82.471\%\\
7y & 235 & 3.551\% & 75.375\%\\
10y & 235 & 3.658\% & 66.998\%\\
\hline
\end{tabular}
\end{center}
\caption{\small Outputs of the calibration on September 10th,
2003. Here $\beta=0.5$ and $H=0.8977$.}\label{output10sep}
\end{table}

\item November 28th, 2003.\\
Stock Price $= 2.297$ euros,  volatility $=14\%$, recovery rate
$=40\%$. The barrier parameter is $H=0.8052$ (re-scaled $V_0=1$).

\begin{table}[h!]
\begin{center}
\begin{tabular}{|c|c|c|c|}
\hline
CDS maturity & Rate (bps) & Volatility nodes & Survival\\
\hline 28-nov-03 & - & 14.081\% & 100.000\%\\
1y & 725 & 14.081\% & 87.620\%\\
3y & 630 & 10.800\% & 72.529\%\\
5y & 570 & 11.489\% & 62.561\%\\
7y & 570 & 16.235\% & 51.810\%\\
10y & 570 & 22.793\% & 39.217\%\\
\hline
\end{tabular}
\end{center}
\caption{\small Outputs of the calibration on November 28th, 2003.
Here $\beta=0.5$ and $H=0.8052$.}\label{output28nov}
\end{table}

\item December 8th, 2003.\\
Stock Price $= 2.237$ euros,  volatility $=20\%$, recovery rate
$=25\%$. The barrier parameter is $H=0.7730$ (re-scaled $V_0=1$).

\begin{table}[h!]
\begin{center}
\begin{tabular}{|c|c|c|c|}
\hline
CDS maturity & Rate (bps) & Volatility nodes & Survival\\
\hline 8-dec-03 & - & 20.197\% & 100.000\%\\
1y & 1450 & 20.197\% & 81.289\%\\
3y & 1200 & 17.972\% & 62.065\%\\
5y & 940 & 13.685\% & 56.263\%\\
7y & 850 & 18.771\% & 49.386\%\\
10y & 850 & 36.661\% & 35.626\%\\
\hline
\end{tabular}
\end{center}
\caption{\small Outputs of the calibration on December 8th, 2003.
Here $\beta=0.5$ and $H=0.7730$.}\label{output8dec}
\end{table}

\item December 10th, 2003.\\
Stock Price $= 2.237$ euros, volatility $=50\%$, recovery rate
$=15\%$. The barrier parameter is $H=0.7253$ (re-scaled $V_0=1$).
\begin{table}[!h]
\begin{center}
\begin{tabular}{|c|c|c|c|}
\hline
CDS maturity & Rate (bps) & Volatility nodes & Survival\\
\hline 10-dec-03 & - & 50.000\% & 100.000\%\\
1y & 5050 & 50.000\% & 55.452\%\\
3y & 2100 & 4.325\% & 55.208\%\\
5y & 1500 & 19.950\% & 50.910\%\\
7y & 1250 & 24.063\% & 46.705\%\\
10y & 1100 & 37.422\% & 39.121\%\\
\hline
\end{tabular}
\end{center}
\caption{\small Outputs of the calibration on December 10th, 2003.
Here $\beta=0.5$ and $H=0.7253$.}\label{output10dec}
\end{table}

\begin{remark} In this last case the calibration algorithm for
the deterministic intensity model could not achieve a solution, so
it was not possible to extract the suitable value for the ``credit
spread based" $H$. In such cases we can proceed in an alternative
way. Let $\bar{\sigma}$ be the equity volatility for the first
maturity that is given by the equity market.
\begin{itemize} \item[a.] Choose a first guess $H^{(1)}$ for $H$. Set i=1;

\item[b.] Calibrate the structural model to the first maturity CDS
market quote through the parameter $\sigma$ when the barrier is
set to $H^{(i)}$; Let $\sigma^{(i)}$ be the calibrated volatility.

\item[c.] Compute the first maturity default probability with the
structural model characterized by $H^{(i)}$ and $\sigma^{(i)}$.

\item[d.] Solve in $H$ the equation where we equate the default
probability computed in ``c" above to the default probability of a
structural model characterized by $H$ and $\bar{\sigma}$; Set
$H^{(i+1)}$ to the found value of $H$; Replace $i$ by $i+1$ and
restart from ``b".

\end{itemize}

Typically, after a number of iterations, $\sigma^{(i)}$ will be
very close to $\bar{\sigma}$ and we may set $H$ to the
corresponding value $H^{(i)}$.
\end{remark}

\end{itemize}

\subsection{Excursion/protection analogy barrier parameter $H$}
Here we present the results of the calibration performed using the
excursion protection analogy to estimate $H$, i.e. $H/V_0 = \rec$.
Now $\beta=0.08$ and the other parameters are the same as before.
\begin{itemize}
\item September 10th, 2003.\\
The barrier parameter is $H=\rec=0.4$ (re-scaled $V_0=1$).
\begin{table}[h!]
\begin{center}
\begin{tabular}{|c|c|c|c|}
\hline
CDS maturity & Rate (bps) & Volatility nodes & Survival\\
\hline 10-sep-03 & - & 42.713\% & 100.000\%\\
1y & 192.5 & 42.713\% & 96.674\%\\
3y & 215 & 26.588\% & 89.524\%\\
5y & 225 & 27.343\% & 82.473\%\\
7y & 235 & 30.686\% & 75.377\%\\
10y & 235 & 31.778\% & 67.002\%\\
\hline
\end{tabular}
\end{center}
\caption{\small Outputs of the calibration on September 10th,
2003.  Here $\beta=0.08$ and $H=0.4$.}\label{output10sep_bis}
\end{table}

\item November 28th, 2003.\\
The barrier parameter is $H=\rec=0.4$ (re-scaled $V_0=1$).
\begin{table}[h!]
\begin{center}
\begin{tabular}{|c|c|c|c|}
\hline
CDS maturity & Rate (bps) & Volatility nodes & Survival\\
\hline 28-nov-03 & - & 58.808\% & 100.000\%\\
1y & 725 & 58.808\% & 87.613\%\\
3y & 630 & 44.225\% & 72.512\%\\
5y & 570 & 46.237\% & 62.541\%\\
7y & 570 & 63.898\% & 51.780\%\\
10y & 570 & 85.758\% & 39.146\%\\
\hline
\end{tabular}
\end{center}
\caption{\small Outputs of the calibration on November 28th, 2003.
Here $\beta=0.08$ and $H=0.4$.}\label{output28nov_bis}
\end{table}

\newpage
\item December 8th, 2003.\\
The barrier parameter is $H=\rec=0.25$ (re-scaled $V_0=1$).
\begin{table}[h!]
\begin{center}
\begin{tabular}{|c|c|c|c|}
\hline
CDS maturity & Rate (bps) & Volatility nodes & Survival\\
\hline 8-dec-03 & - & 107.846\% & 100.000\%\\
1y & 1450 & 107.846\% & 81.277\%\\
3y & 1200 & 94.506\% & 62.041\%\\
5y & 940 & 71.046\% & 56.246\%\\
7y & 850 & 96.431\% & 49.369\%\\
10y & 850 & 181.575\% & 35.559\%\\
\hline
\end{tabular}
\end{center}
\caption{\small Outputs of the calibration on December 8th, 2003.
Here $\beta=0.08$ and $H=0.25$.}\label{output8dec_bis}
\end{table}

\item December 10th, 2003.\\
The barrier parameter is $H=\rec=0.15$ (re-scaled $V_0=1$).

\begin{table}[!h]
\begin{center}
\begin{tabular}{|c|c|c|c|}
\hline
CDS maturity & Rate (bps) & Volatility nodes & Survival\\
\hline 10-dec-03 & - & 292.060\% & 100.000\%\\
1y & 5050 & 292.060\% & 55.381\%\\
3y & 2100 & 21.603\% & 55.198\%\\
5y & 1500 & 114.713\% & 50.904\%\\
7y & 1250 & 137.488\% & 46.700\%\\
10y & 1100 & 210.474\% & 39.104\%\\
\hline
\end{tabular}
\end{center}
\caption{\small Outputs of the calibration on December 10th, 2003.
Here $\beta=0.08$ and $H=0.15$.}\label{output10dec_bis}
\end{table}
\end{itemize}
Finally, we present the calibration performed for the first date
(September 10th, 2003) with $\beta=0.08$ and $H$ estimated by
means of the credit spread method. Comparing the values in Table
\ref{output10sep_ter} with those in Table \ref{output10sep} we see
that the volatilities are nearly the same, and this holds in
general, at least when the volatilities are not too high. Here the
barrier parameter is $H=0.8969$ (re-scaled $V_0=1$).
\begin{table}[!h]
\begin{center}
\begin{tabular}{|c|c|c|c|}
\hline
CDS maturity & Rate (bps) & Volatility nodes & Survival\\
\hline 10-sep-03 & - & 5.012\% & 100.000\%\\
1y & 192.5 & 5.012\% & 96.673\%\\
3y & 215 & 3.064\% & 89.522\%\\
5y & 225 & 3.110\% & 82.468\%\\
7y & 235 & 3.445\% & 75.369\%\\
10y & 235 & 3.508\% & 66.988\%\\
\hline
\end{tabular}
\end{center}
\caption{\small Outputs of the calibration on September 10th,
2003. Here $\beta=0.08$ and $H=0.8969$.}\label{output10sep_ter}
\end{table}

\subsection{Discount Curves}
In the following we report the Discount Curves that have been used
in the simulations.
\begin{itemize}
\item September 10th, 2003.\\
\vspace{-0.5cm}
{\small
\begin{table}[h!]
\begin{center}
\begin{tabular}{|c|c||c|c||c|c|}
\hline
Date &   Discount & Date &   Discount & Date &   Discount \\
\hline
Sep 10th, 03 &    1.00000 & Jun 14th, 04 &    0.98320 & Sep 12th, 11 &    0.71837 \\
Sep 11th, 03 &    0.99994 & Sep 13th, 04 &    0.97721 & Sep 12th, 12 &    0.68171 \\
Sep 15th, 03 &    0.99971 & Sep 12th, 05 &    0.94740 & Sep 12th, 13 &    0.64624 \\
Sep 19th, 03 &    0.99948 & Sep 12th, 06 &    0.91195 & Sep 12th, 14 &    0.61204 \\
Oct 13th, 03 &    0.99805 & Sep 12th, 07 &    0.87373 & Sep 14th, 15 &    0.58032 \\
Nov 12th, 03 &    0.99627 & Sep 12th, 08 &    0.83501 & Sep 12th, 16 &    0.54942 \\
Dec 12th, 03 &    0.99447 & Sep 14th, 09 &    0.79557 & Sep 12th, 17 &    0.51882 \\
Mar 12th, 04 &    0.98896 & Sep 13th, 10 &    0.75666 & Sep 12th, 18 &    0.48937 \\
\hline
\end{tabular}
\caption{\small Discount curve on September 10th, 2003.}
\end{center}
\end{table}}

\item November 28th, 2003.\\
\vspace{-0.5cm}
{\small
\begin{table}[h!]
\begin{center}
\begin{tabular}{|c|c||c|c||c|c|}
\hline
Date &   Discount & Date &   Discount & Date &   Discount \\
\hline
Nov 28th, 03 &    1.00000 & Sep 02nd, 04 &    0.98200 & Dec 02nd, 11 &    0.70648 \\
Dec 01st, 03 &    0.99983 & Dec 02nd, 04 &    0.97508 & Dec 03rd, 12 &    0.66950 \\
Dec 03rd, 03 &    0.99971 & Dec 02nd, 05 &    0.94202 & Dec 02nd, 13 &    0.63449 \\
Dec 09th, 03 &    0.99936 & Dec 04th, 06 &    0.90424 & Dec 02nd, 14 &    0.60081 \\
Jan 02nd, 04 &    0.99792 & Dec 03rd, 07 &    0.86449 & Dec 02nd, 15 &    0.56852 \\
Feb 02nd, 04 &    0.99607 & Dec 02nd, 08 &    0.82431 & Dec 02nd, 16 &    0.53760 \\
Mar 02nd, 04 &    0.99435 & Dec 02nd, 09 &    0.78427 & Dec 04th, 17 &    0.50786 \\
Jun 02nd, 04 &    0.98849 & Dec 02nd, 10 &    0.74490 & Dec 03rd, 18 &    0.47965 \\
\hline
\end{tabular}
\caption{\small Discount curve on November 28th, 2003.}
\end{center}
\end{table}}

\item December 8th, 2003.\\
\vspace{-0.5cm}
{\small
\begin{table}[h!]
\begin{center}
\begin{tabular}{|c|c||c|c||c|c|}
\hline
Date &   Discount & Date &   Discount & Date &   Discount \\
\hline
Dec 08th, 03 &    1.00000 & Sep 10th, 04 &    0.98271 & Dec 12th, 11 &    0.71134 \\
Dec 09th, 03 &    0.99994 & Dec 10th, 04 &    0.97612 & Dec 10th, 12 &    0.67460 \\
Dec 11th, 03 &    0.99983 & Dec 12th, 05 &    0.94456 & Dec 10th, 13 &    0.63960 \\
Dec 17th, 03 &    0.99949 & Dec 11th, 06 &    0.90787 & Dec 10th, 14 &    0.60587 \\
Jan 12th, 04 &    0.99793 & Dec 10th, 07 &    0.86880 & Dec 10th, 15 &    0.57341 \\
Feb 10th, 04 &    0.99620 & Dec 10th, 08 &    0.82911 & Dec 12th, 16 &    0.54209 \\
Mar 10th, 04 &    0.99448 & Dec 10th, 09 &    0.78937 & Dec 11th, 17 &    0.51229 \\
Jun 10th, 04 &    0.98882 & Dec 10th, 10 &    0.74996 & Dec 10th, 18 &    0.48373 \\
\hline
\end{tabular}
\caption{\small Discount curve on December 8th, 2003.}
\end{center}
\end{table}}

\item December 10th, 2003.\\
\vspace{-0.5cm}
{\small
\begin{table}[h!]
\begin{center}
\begin{tabular}{|c|c||c|c||c|c|}
\hline
Date &   Discount & Date &   Discount & Date &   Discount \\
\hline
Dec 10th, 03 &    1.00000 & Sep 13th, 04 &    0.98268 & Dec 12th, 11 &    0.71165 \\
Dec 11th, 03 &    0.99994 & Dec 13th, 04 &    0.97610 & Dec 12th, 12 &    0.67476 \\
Dec 15th, 03 &    0.99972 & Dec 12th, 05 &    0.94476 & Dec 12th, 13 &    0.63957 \\
Dec 19th, 03 &    0.99949 & Dec 12th, 06 &    0.90802 & Dec 12th, 14 &    0.60571 \\
Jan 12th, 04 &    0.99805 & Dec 12th, 07 &    0.86889 & Dec 14th, 15 &    0.57303 \\
Feb 12th, 04 &    0.99621 & Dec 12th, 08 &    0.82922 & Dec 12th, 16 &    0.54191 \\
Mar 12th, 04 &    0.99448 & Dec 14th, 09 &    0.78919 & Dec 12th, 17 &    0.51196 \\
Jun 14th, 04 &    0.98874 & Dec 13th, 10 &    0.74987 & Dec 12th, 18 &    0.48324 \\
\hline
\end{tabular}
\end{center}
\caption{\small Discount curve on December 10th, 2003.}
\end{table}}
\item March 10th, 2004.\\
\vspace{-0.5cm}
{\small
\begin{table}[h!]
\begin{center}
\begin{tabular}{|c|c||c|c||c|c|}
\hline Date &   Discount & Date &   Discount & Date &   Discount\\
\hline
Mar 10th, 04 &    1.00000 & Dec 13th, 04 &    0.98422 & Mar 12th, 12 &    0.73537\\
Mar 11th, 04 &    0.99994 & Mar 14th, 05 &    0.97884 & Mar 12th, 13 &    0.69831\\
Mar 15th, 04 &    0.99972 & Mar 13th, 06 &    0.95375 & Mar 12th, 14 &    0.66271\\
Mar 19th, 04 &    0.99949 & Mar 12th, 07 &    0.92279 & Mar 12th, 15 &    0.62817\\
Apr 13th, 04 &    0.99807 & Mar 12th, 08 &    0.88770 & Mar 14th, 16 &    0.59552\\
May 12th, 04 &    0.99642 & Mar 12th, 09 &    0.85063 & Mar 13th, 17 &    0.56392\\
Jun 14th, 04 &    0.99455 & Mar 12th, 10 &    0.81234 & Mar 12th, 18 &    0.53344\\
Sep 13th, 04 &    0.98940 & Mar 14th, 11 &    0.77355 & Mar 12th, 19 &    0.50440\\
\hline
\end{tabular}
\end{center}
\caption{\small Discount curve on March 10th, 2004.}
\end{table}}
\end{itemize}

\end{document}